\documentclass[journal]{IEEEtran}
\IEEEoverridecommandlockouts
\newtheorem{Theorem}{Theorem}

\newtheorem{Corollary}{Corollary}

\newtheorem{Remark}{Remark}
\usepackage{algorithm}
\usepackage{algorithmic}
\usepackage{graphicx}
\usepackage{textcomp}
\usepackage{stfloats}
\usepackage{subfigure}
\usepackage{amssymb,amsmath,bm}
\usepackage{cite}
\usepackage{times}
\usepackage{latexsym}
\usepackage{array}
\usepackage{fancyhdr}
\usepackage{psfrag}
\usepackage{multirow}
\usepackage{xcolor}
\usepackage{color}
\usepackage{makecell}
\usepackage{threeparttable}
\usepackage{hyperref} 
\usepackage[OT1]{fontenc}

\begin{document} 

\title{Channel Estimation for Double-BD-RIS-Assisted Multi-User MIMO Communication\\}

 \author{\IEEEauthorblockN{Junyuan~Gao, 
 Shuowen~Zhang, 
 and Liang~Liu 
 }
 \thanks{J. Gao, S. Zhang, and L. Liu are with the Department of Electrical and Electronic Engineering, The Hong Kong Polytechnic University, Hong Kong SAR (e-mail: \{junyuan.gao, shuowen.zhang, liang-eie.liu\}@polyu.edu.hk).} 
 }

 \maketitle

\begin{abstract}

Deploying multiple beyond diagonal reconfigurable intelligent surfaces~(BD-RISs) can potentially improve the communication performance thanks to inter-element connections of each BD-RIS and inter-surface cooperative beamforming gain among BD-RISs. However, a major issue for multi-BD-RIS-assisted communication lies in the channel estimation overhead - the channel coefficients associated with the off-diagonal elements in each BD-RIS's scattering matrix as well as those associated with the reflection links among BD-RISs have to be estimated. In this paper, we propose an efficient channel estimation framework for double-BD-RIS-assisted multi-user multiple-input multiple-output~(MIMO) systems. Specifically, we reveal that high-dimensional cascaded channels are characterized by five low-dimensional matrices by exploiting channel correlation properties. Based on this novel observation, in the ideal noiseless case, we develop a channel estimation scheme to recover these matrices sequentially and characterize the closed-form overhead required for perfect estimation as a function of the numbers of users and each BD-RIS's elements and channel ranks, which is with the same order as that in double-diagonal-RIS-aided communication systems. This exciting result implies the superiority of cooperative BD-RIS-aided communication over the diagonal-RIS counterpart even when channel estimation overhead is considered. We further extend the proposed scheme to practical noisy scenarios and provide extensive numerical simulations to validate its effectiveness. 

\end{abstract}

\section{Introduction}\label{sec:intro} 

\subsection{Motivation}

Reconfigurable intelligent surface (RIS) has been recognized as a promising technique in future sixth-generation~(6G) cellular networks. It is able to 
control the radio propagation environment in favor of signal transmission, thereby enhancing the wireless communication performance~\cite{ref:RIS_QingqingWu,ref:RIS_ZhangSW}. 
In conventional RIS-aided communication, each reflecting element is adjusted independently such that the scattering matrix is diagonal. 
However, this diagonal architecture limits the flexibility in wave manipulation. 
To address this issue, a new technology, known as the beyond diagonal RIS (BD-RIS), has been proposed and attracted significant attention~\cite{ref:BDRIS_benefit1,ref:BDRIS_benefit2}.   
By introducing interconnections among reflecting elements, the scattering matrix of BD-RIS exhibits a more general non-diagonal structure, thereby offering greater flexibility in wave manipulation and leading to enhanced beamforming gain, spectral efficiency, and coverage compared to traditional diagonal-RIS-aided communication. 
Given the appealing potential of BD-RIS, extensive interests have been devoted to this area, including modeling and architecture design~\cite{ref:BDRIS_benefit2,ref:BDRIS_mode}, performance optimization and beamforming design~\cite{ref:BDRIS_optimization1,ref:BDRIS_optimization2}, channel estimation~\cite{ref:wangrui,ref:BDRIS_CE_LS,ref:BDRIS_CE_tensor}, etc.

\begin{figure}
    \centering
    \includegraphics[width=0.95\linewidth]{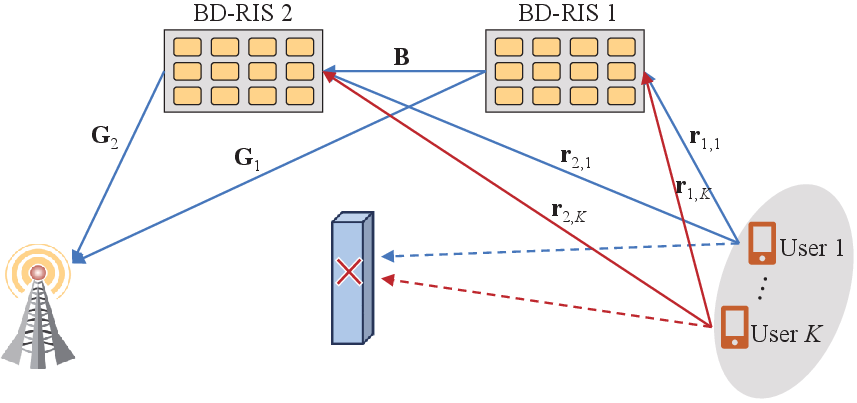}\\
    \caption{System model of double-BD-RIS-aided communication. BD-RIS 1 is close to the users, and BD-RIS 2 is close to the BS. The BS receives signals via the user - BD-RIS 1 - BS link, user - BD-RIS 2 - BS link, and user - BD-RIS 1 - BD-RIS 2 - BS link.}
  \label{fig:SM} 
  \end{figure}

A common assumption in existing BD-RIS research is that merely one BD-RIS exists in the network. However, in practice, multiple BD-RISs will co-exist to serve the distributed users, and a joint design of multiple BD-RISs' strategies is necessary by considering the impact of inter-BD-RIS channels. 
In the literature, plenty of works have been done to study the issues about channel estimation and joint scattering matrix design in multi-diagonal-RIS-aided communication \cite{ref:doubleRIS_Shuowen,ref:MultiRIS,ref:doubleRIS_ZhengBY_TWC,ref:doubleRIS_YouCS,ref:doubleRIS_ZhengBY_conf,ref:doubleRIS_ZhengBY_TCOM}. However, to the best of our knowledge, the investigation of multi-BD-RIS-aided communication is still missing. To bridge this gap, in this paper, we will start with a double-BD-RIS-assisted multi-user communication system with the co-existence of both single- and double-reflection links, as shown in Fig. \ref{fig:SM}.

Exploiting the full potential of double-BD-RIS architectures relies heavily on the availability of accurate channel state information (CSI).  
However, acquiring CSI of double-BD-RIS-aided systems is particularly challenging due to the non-diagonal structure of BD-RIS scattering matrices and the co-existence of single- and double-reflection links, which result in 
a significantly large number of channel coefficients to be estimated. 
Motivated by this, for the first time in the literature, this work aims to explore channel estimation strategies for double-BD-RIS systems. 
In particular, we will demonstrate that the training overhead of the double-BD-RIS system can be reduced to the same order as that in both single-BD-RIS and double-diagonal-RIS counterparts. 

\subsection{Prior Works}

In the scenario with a single diagonal RIS, classical methods, such as least squares (LS) and linear minimum mean-squared error (LMMSE), can be applied for channel estimation but at the cost of high training overheads \cite{ref:RIS_Mag}. Consequently, a variety of approaches were proposed to reduce the overhead, 
such as grouping RIS elements~\cite{ref:RIS_YouCS,ref:RIS_ZhengBX} and leveraging geometric models and channel sparsity~\cite{ref:RIS_ZhouG,ref:RIS_He}. 
Moreover, the multi-user channel correlation property was exploited in \cite{ref:WangZR,ref:wangrui_feedback} to achieve low-overhead channel estimation. It was proved that in the scenario with $K$ single-antenna users, an $L$-antenna BS, and an $M_1$-element diagonal RIS, it is sufficient to utilize $M_1 +\left\lceil \frac{(K-1)M_1}{q_1} \right\rceil$ pilot symbols to perfectly recover the cascaded channels in the noiseless case, where $q_1$ denotes the rank of the BS-RIS channel \cite{ref:WangZR}.

Compared to diagonal-RIS architectures, it is more challenging to perform channel estimation for BD-RIS-aided systems, since a large number of channel coefficients associated with off-diagonal entries of the scattering matrix have to be estimated as well. 
The authors in \cite{ref:BDRIS_CE_LS} proposed a closed-form solution to design the BD-RIS scattering matrix based on the LS criterion, which, however, suffers from extremely high overhead that scales prohibitively with the circuit complexity of the BD-RIS.  
To address this, \cite{ref:BDRIS_CE_tensor} showed that the overhead can be reduced by exploiting the built-in block Kronecker structure of the composite channel. 
In \cite{ref:wangrui}, the authors demonstrated that by exploiting the fact that for any given user antenna, its cascaded channel matrix associated with one reference BD-RIS element is a scaled version of that associated with any other element, the training overhead in the BD-RIS-aided network can be further reduced to $2M_1 + \left\lceil \tfrac{M_1(K-1)}{q_1} \right\rceil$, which is on the same order as that required in the conventional diagonal RIS network.


Motivated by the appealing cooperative gain of double-RIS systems over single-RIS counterparts, existing literature has explored the channel estimation problem for double-RIS-aided networks, but concentrating on the simple double-diagonal-RIS case. 
Specifically, assuming that a single-antenna user communicates with a single-antenna BS via the double-reflection link, the LS criterion was utilized in \cite{ref:doubleRIS_YouCS} to estimate the cascaded channels. 
The authors in \cite{ref:doubleRIS_ZhengBY_conf} considered a double-diagonal-RIS-aided multi-user communication system where each user is equipped with a single antenna and the BS is equipped with multiple antennas, and presented a decoupled channel estimation scheme, which was shown to exhibit much lower overhead than the LS estimator. 
Particularly, under the assumptions that the numbers of elements of two RISs are on the same order and the rank of the channel between the BS and each RIS is dominated by the number of RIS elements, the minimum training overhead of the double-diagonal-RIS-aided system is on the same order as that of the single-diagonal-RIS counterpart \cite{ref:doubleRIS_Mag}.


Despite existing research, achieving high-accuracy channel estimation for double-BD-RIS-aided systems with low overhead remains an essential yet challenging open problem. 
The main difficulties are as follows. 
First, the number of channel coefficients to be estimated in double-BD-RIS-aided systems is prohibitive. Specifically, compared to the traditional diagonal-RIS-aided network, the inter-connected architecture of BD-RIS necessitates the estimation of more coefficients associated with off-diagonal elements of the scattering matrix. 
Compared to the single-BD-RIS network, which only consists of single-reflection links, the double-BD-RIS architecture introduces an additional double-reflection link that induces a substantial increase in the dimensionality of the channel matrix. 
Second, the channels in double-BD-RIS systems are intricately coupled. This arises from the overlap in propagation environments between the single- and double-reflection links, as well as the non-diagonal structure of the BD-RIS scattering matrix that induces signal mixing across elements. 
Third, in double-BD-RIS systems, the scattering matrices of two BD-RISs should be cooperatively designed. Due to circuit constraints, each BD-RIS scattering matrix is unitary, implying that all elements of a scattering matrix should be adjusted jointly~\cite{ref:BDRIS_benefit2,ref:BDRIS_mode}. The necessity to cooperatively design two unitary matrices complicates the time-varying reflection pattern design for channel estimation. 
These challenges raise a critical question: How can we achieve high-accuracy channel estimation for double-BD-RIS-aided systems with low overhead, and more importantly, is it possible to reduce this overhead to the same order as that of the single-BD-RIS or double-diagonal-RIS case? This work aims to provide an affirmative answer to this question by developing an efficient channel estimation strategy that fully exploits channel correlation properties.


%
%


\subsection{Main Contributions}


In this paper, we consider a double-BD-RIS-assisted multi-user multiple-input multiple-output (MIMO) communication system in the uplink, as shown in Fig.~\ref{fig:SM}. The first BD-RIS with $M_1$ elements is deployed near a cluster of $K$ single-antenna users and the second one with $M_2$ elements is deployed near the $L$-antenna BS. The pilots transmitted by users reach the BS via the single- and double-reflection links. 
Under this setup, our objective is to estimate the cascaded channels by jointly designing the scattering matrices of two BD-RISs and the pilot sequences of $K$ users.  
The main contributions of this paper are summarized as follows.

\begin{itemize}  
    \item  
    We leverage intrinsic channel correlation properties in double-BD-RIS-aided multi-user MIMO networks to substantially reduce the number of elements to be estimated. 
    Specifically, for the single-reflection link via BD-RIS $i \in \{1,2\}$, the cascaded channel corresponding to any user and incident element of BD-RIS $i$ is a scaled version of the linear combination of the channels from all users to an element of BD-RIS $i$ and then to the BS. 
    For the double-reflection link, the cascaded channel associated with any user $k$, element pair $(m_1, m'_1)$ of BD-RIS 1 (denoting the $m_1$-th incident and $m'_1$-th reflecting elements), and the $m_2$-th incident element of BD-RIS 2 is also a scaled version of a reference channel. 
    Moreover, the propagation environment of each single-reflection link and the double-reflection link is overlapped.     
    Based on these properties, we reveal that the high-dimensional cascaded channels can be fully characterized by five matrices of much lower dimensions.

    
    

    \item  
    In the noiseless case, we design a low-overhead channel estimation framework for double-BD-RIS networks, 
    which consists of five phases to estimate the five low-dimensional matrices sequentially. 
    Further, we show that under our approach, a time duration consisting of 
    $4(M_1 + M_2) + 2  \left\lceil \tfrac{KM_2}{q_2} \right\rceil  +  2 \left\lceil \tfrac{M_1M_2}{q_2}  \right\rceil +  \left\lceil \tfrac{KM_1}{ f }  \right\rceil $ 
    pilot symbols is sufficient for perfect channel estimation, where $q_2$ denotes the rank of the BD-RIS 2 - BS channel, and $f$ is related to the rank of the channel aggregated by BD-RIS 1 - BS and BD-RIS 1 - BD-RIS 2 - BS links. 
    This overhead is of a much smaller order than that required by treating all entries of the cascaded channels as independent elements. 
    More importantly, this overhead is with the same order as that for the single-diagonal-RIS-aided communication \cite{ref:WangZR}, that for the double-diagonal-RIS-aided communication \cite{ref:doubleRIS_ZhengBY_conf}, as well as that for the single-BD-RIS-aided communication \cite{ref:wangrui} under mild propagation conditions. This indicates the theoretical performance gain of multi-BD-RIS-assisted communication over the other three counterparts even with channel estimation overhead considered. 

    \item  
    We extend the proposed channel estimation framework to the practical case with noise at the BS, which is applicable in the absence of any prior knowledge on the cascaded channels and does not rely on the ON/OFF reflection control of BD-RISs.
    We provide extensive numerical results to validate its effectiveness. 
    Specifically, the proposed scheme demonstrates a significant performance advantage over the benchmark scheme that treats all entries of the cascaded channels as independent elements. 
    Moreover, by allowing all users to transmit pilots simultaneously, the proposed scheme achieves superior channel estimation performance than applying the idea in \cite{ref:wangrui,ref:doubleRIS_ZhengBY_conf}, where $K-1$ users are required to remain silent during the estimation of the reference channel.

\end{itemize}

%
%
%
%

\emph{Organizations:} The rest of this paper is organized as follows. Section~\ref{Sec:model} describes the double-BD-RIS-aided system model. 
Section~\ref{Sec:III} reveals the channel property and formulates the channel estimation problem. 
In Section~\ref{Sec:CE_noAWGN}, we propose a channel estimation framework for the double-BD-RIS-aided network and characterize the overhead for perfectly estimating the channels in the noiseless case. 
Section~\ref{Sec:CE_AWGN} extends the channel estimation strategy to a general case with noise at the BS.
Numerical results are presented in Section~\ref{Sec:simulation}.  
Finally, Section~\ref{Sec:conclusion} concludes this paper.

  \emph{Notations:} Throughout this paper, uppercase and lowercase boldface letters denote matrices and vectors, respectively.
  The notations $\left[\mathbf{X} \right]_{m,n}$, $\left[\mathbf{X} \right]_{m,:}$, and $\left[\mathbf{X} \right]_{:,n}$ denote the $\left( m,n \right)$-th element, the $m$-th row, and the $n$-th column of the matrix $\mathbf{X}$, respectively. 
  We use $\operatorname{Row}(\mathbf{X})$, $\operatorname{Col}(\mathbf{X})$, $\mathbf{X}^{-1}$, and $\operatorname{rank}(\mathbf{X})$ to denote the row space, column space, inverse, and rank of the matrix $\mathbf{X}$, respectively. 
  We use $\left(\cdot \right)^{*}$, $\left(\cdot \right)^{T}$, $\left(\cdot \right)^{H}$, $\left\|\mathbf{x} \right\|_{p}$, and $\left\|\mathbf{X} \right\|_{F}$ to denote conjugate, transpose, 
  conjugate transpose, ${\ell}_p$-norm, and Frobenius norm, respectively. 
  The notation $\mathbf{I}_{n}$ denotes an $n\times n$ identity matrix. 
  The notation $\operatorname{arg}(\cdot)$ returns the angle of a complex number. 
  The notation $\dim(\cdot)$ returns the dimension of a space. 
  The notations $\lfloor \cdot \rfloor$, $\lceil \cdot \rceil$, $\otimes$, $\cdot\backslash\cdot$, and $\operatorname{mod}$ denote the floor function, ceiling function, Kronecker product, set subtraction, and modulo operation, respectively.  
  We use $\operatorname{diag} \left\{ \mathbf{x} \right\}$ to denote a diagonal matrix with $\mathbf{x}$ comprising its diagonal elements, and use $\operatorname{blkdiag} \left\{ \mathbf{A}, \mathbf{B} \right\}$ to denote a block diagonal matrix. 
  The notation $\operatorname{vec}(\cdot)$ denotes the vectorization of a matrix, and $\operatorname{unvec}(\cdot)$ denotes its reverse operation. 
  For an integer $k > 0$, let $[k] = \left\{1,\ldots,k \right\}$. For integers $k_2 \geq k_1$, let $[k_1,k_2] = \left\{k_1,\ldots,k_2\right\}$. 
  We use $1[\cdot]$ to denote the indicator function. 
  The distribution of a circularly symmetric complex Gaussian~(CSCG) random vector $\mathbf{x}$ with mean $\bm{\mu}$ and covariance matrix $\bm{\Sigma}$ is denoted as $\mathcal{CN}(  \bm{\mu} , \bm{\Sigma} )$.  
  Let $f(x)$ and $g(x)$ be positive. The notation $f(x) = \Theta \left( g(x)\right)$ means that $\lim_{x\to\infty} f(x)/g(x)$ is equal to a positive constant.

\section{System Model}\label{Sec:model}

We consider a double-BD-RIS-aided uplink communication system as illustrated in Fig. \ref{fig:SM}, where two BD-RISs, referred to as BD-RIS 1 and BD-RIS 2, are respectively deployed near a cluster of $K$ single-antenna users and near an $L$-antenna BS to assist their communication. 
We assume that both BD-RIS $1$ and BD-RIS $2$ adopt a fully connected architecture, comprising $M_1$ and $M_2$ passive reflecting elements, respectively. 
Due to the circuit requirement, the scattering matrix of BD-RIS $i\in\{1,2\}$ at time instant $t$, denoted as $\bm{\Phi}_{i,t} \in \mathbb{C}^{M_i \times M_i}$, is a unitary matrix satisfying~\cite{ref:BDRIS_benefit2,ref:BDRIS_mode} 
\begin{equation} \label{eq:Phi}
  \bm{\Phi}^H_{i,t} \bm{\Phi}_{i,t} = \bm{\Phi}_{i,t} \bm{\Phi}^H_{i,t} = \mathbf{I}_{M_i},  \;\; \forall i \in \{1,2\},t. 
\end{equation} 

We assume that the direct links between the users and the BS are blocked due to obstacles~\cite{ref:doubleRIS_ZhengBY_TCOM,ref:doubleRIS_Mag}. Thus, the communication between them relies on the single reflection links, i.e., the user $k$ - BD-RIS $i$ - BS links, $\forall k \in [K], i \in \{1,2\}$, and the double reflection links, i.e., the user $k$ - BD-RIS $1$ - BD-RIS $2$ - BS links, $\forall k \in [K]$.\footnote{The user - BD-RIS 2 - BD-RIS 1 - BS links are too weak compared to other links because BD-RIS 1 is near the users and BD-RIS 2 is near the BS. Therefore, similar to \cite{ref:doubleRIS_Mag,ref:doubleRIS_ZhengBY_TCOM,ref:doubleRIS_ZhengBY_TWC}, this channel is ignored in this paper.}
We consider a quasi-static block fading channel model, where the channel remains constant in each coherence block. 
Specifically, the channel from user $k$ to the $m_i$-th element of BD-RIS $i$ is denoted as $r_{i,k,m_i}$, the channel from the $m_1$-th element of BD-RIS $1$ to the $m_2$-th element of BD-RIS $2$ is denoted as $b_{m_2,m_1}$, and the channel from the $m_i$-th element of BD-RIS $i$ to the BS is denoted as $\mathbf{g}_{i,m_i} \in \mathbb{C}^{L}$, $\forall i\in\{1,2\}, k\in[K], m_1\in[M_1],$ and $m_2\in[M_2]$. 
Then, the overall channel from user $k$ to BD-RIS $i$ is defined as $\mathbf{r}_{i,k} = [r_{i,k,1}, \ldots, r_{i,k,M_i}]^T \in \mathbb{C}^{M_i}$, $\forall i\in\{1,2\}, k \in [K]$. 
Denote $\mathbf{R}_i = [\mathbf{r}_{i,1}, \ldots, \mathbf{r}_{i,K} ] \in \mathbb{C}^{M_i\times K}, \forall i\in\{1,2\}$. 
The overall channel between BD-RIS $1$ and BD-RIS $2$ is characterized by the matrix $\mathbf{B}\in \mathbb{C}^{M_2\times M_1}$, with its $(m_2,m_1)$-th element given by $b_{m_2,m_1}$. 
The overall channel from BD-RIS $i$ to the BS is denoted as $\mathbf{G}_i = [\mathbf{g}_{i,1}, \ldots, \mathbf{g}_{i,M_i}] \in \mathbb{C}^{L\times M_i}, \forall i\in\{1,2\}$. 
Then, at time instant $t$, the channel of the single-reflection link, i.e., user $k$ - BD-RIS $i$ - BS channel, is given by 
\begin{equation} \label{eq:h1kt} 
  \mathbf{h}_{i,k,t} = \mathbf{G}_i \bm{\Phi}_{i,t} \mathbf{r}_{i,k} = \mathbf{J}_{i,k} \operatorname{vec}(\bm{\Phi}_{i,t}) , \;\; \forall i \in \{1,2\}, k,t,  
\end{equation}
where the cascaded channel $\mathbf{J}_{i,k}$ is given by 
\begin{align} 
  \mathbf{J}_{i,k} = \mathbf{r}_{i,k}^T \otimes \mathbf{G}_i & = [ \mathbf{Q}_{i,k,1}, \mathbf{Q}_{i,k,2}, \ldots, \mathbf{Q}_{i,k,M_i} ], \notag\\
  &  \;\;\;\;\;\;\;\;\;\;\;\;\;\;\;\;\;\;\;\;\;\;\;\;\;\; \forall i \in \{1,2\},k, \label{eq:J1k}
\end{align} 
with $\mathbf{Q}_{i,k,m_i} = r_{i,k,m_i} \mathbf{G}_i$ denoting the cascaded channel from user $k$ to the $m_i$-th element of BD-RIS $i$ and then to the BS, $\forall i\in\{1,2\},k\in[K],m_i\in[M_i]$. 
The channel of the double reflection link, i.e., user $k$ - BD-RIS $1$ - BD-RIS $2$ - BS channel, is expressed as
\begin{subequations}
\begin{align}
  \mathbf{h}_{1,2,k,t} & = \mathbf{G}_2 \bm{\Phi}_{2,t} \mathbf{B} \bm{\Phi}_{1,t} \mathbf{r}_{1,k}  \\
  & = \left( \mathbf{r}_{1,k}^T \otimes \mathbf{G}_2 \right) \operatorname{vec}\left( \bm{\Phi}_{2,t} \mathbf{B} \bm{\Phi}_{1,t} \right) \\
  & = \left( \mathbf{r}_{1,k}^T \otimes \mathbf{G}_2 \right) \left( \bm{\Phi}_{1,t}^T \otimes \bm{\Phi}_{2,t} \right) 
  \operatorname{vec}\left( \mathbf{B} \right) \\
  & = \mathbf{J}_{1,2,k} \operatorname{vec}( \bm{\Phi}_{1,t}^T \otimes \bm{\Phi}_{2,t} ) , \;\; \forall k,t, \label{eq:h12kt}
\end{align} 
\end{subequations}
where the cascaded channel $\mathbf{J}_{1,2,k}$ is given by 
\begin{subequations}
\begin{align} 
  \mathbf{J}_{1,2,k} & = \operatorname{vec}^T( \mathbf{B} ) \otimes \mathbf{r}_{1,k}^T \otimes \mathbf{G}_2 \\
  & = \left[ \mathbf{Q}_{1,2,k,1,1,1}, \ldots, \mathbf{Q}_{1,2,k,1,1,M_1}, \ldots , \right. \notag\\
  & \;\;\;\;\; \; \mathbf{Q}_{1,2,k,M_2,1,1}, \ldots, \mathbf{Q}_{1,2,k,M_2,1,M_1}, \ldots,  \notag\\
  & \;\;\;\;\; \; \mathbf{Q}_{1,2,k,1,M_1,1}, \ldots, \mathbf{Q}_{1,2,k,1,M_1,M_1}, \ldots , \notag\\ 
  & \;\;\;\;\; \left. \mathbf{Q}_{1,2,k,M_2,M_1,1}, \ldots, \mathbf{Q}_{1,2,k,M_2,M_1,M_1} \right], \;\;  \forall k, \label{eq:J12k}
\end{align}
\end{subequations}
with $\mathbf{Q}_{1,2,k,m_2,m_1,m'_1} = b_{m_2,m_1} r_{1,k,m'_1} \mathbf{G}_2$ denoting the cascaded channel from user $k$ to the element pair $(m_1, m'_1)$ of BD-RIS 1 (representing the $m_1$-th incident and $m'_1$-th reflecting elements), then to the $m_2$-th incident element of BD-RIS 2, and finally to the BS, $\forall k,m_1,m'_1,m_2$. 

The received signal of the BS at time instant $t$ is given by
\begin{subequations}
\begin{align} 
  \mathbf{y}_t 
  & = {\sum}_{k=1}^{K} \sqrt{p} x_{k,t} ( \mathbf{h}_{1,k,t} + \mathbf{h}_{2,k,t} + \mathbf{h}_{1,2,k,t} ) + \mathbf{z}_t  \\
  & = {\sum}_{k=1}^{K} \sqrt{p} x_{k,t} \left( \mathbf{J}_{1,k} \operatorname{vec}(\bm{\Phi}_{1,t}) + \mathbf{J}_{2,k} \operatorname{vec}(\bm{\Phi}_{2,t}) \right. \notag\\
  & \;\;\;\;\; \left.+ \mathbf{J}_{1,2,k} \operatorname{vec}( \bm{\Phi}_{1,t}^T \otimes \bm{\Phi}_{2,t} ) \right) + \mathbf{z}_t ,   \;\; \forall t, \label{eq:yt}
\end{align}
\end{subequations}
where $p$ denotes the common user transmit power, $x_{k,t}$ denotes the unit-power pilot of user $k$ at time instant $t$, and $\mathbf{z}_t \sim \mathcal{CN}(\bm{0}, \sigma^2 \mathbf{I}_L)$ denotes the additive white Gaussian noise (AWGN) of the BS at time instant $t$.

In this work, we aim to estimate the cascaded channels $\mathbf{J}_{i,k}$ in \eqref{eq:J1k} and $\mathbf{J}_{1,2,k}$ in \eqref{eq:J12k}, $\forall i\in\{1,2\}, k\in[K]$, based on the received signals $\mathbf{y}_t, \forall t$, given in \eqref{eq:yt}. 
Due to inter-element connections of each BD-RIS and the co-existence of single- and double-reflection links, the number of elements in $\mathbf{J}_{1,k}$'s, $\mathbf{J}_{2,k}$'s, and $\mathbf{J}_{1,2,k}$'s reaches $KL(M_1^2 + M_2^2 + M_1^2M_2^2)$, which is significantly larger than the number of channel coefficients in the network with a single $M_1$-element BD-RIS, that of the network with a single $M_1$-element diagonal RIS, and that of the double-diagonal-RIS network, given by $KLM_1^2$ \cite{ref:wangrui}, $KLM_1$ \cite{ref:WangZR}, and $KL(M_1 + M_2 + M_1 M_2)$ \cite{ref:doubleRIS_ZhengBY_conf}, respectively. 


\section{Problem Statement} \label{Sec:III}

It is observed from \eqref{eq:yt} that $\mathbf{y}_t$ is linear with respect to the cascaded channels $\mathbf{J}_{1,k}$'s, $\mathbf{J}_{2,k}$'s, and $\mathbf{J}_{1,2,k}$'s. 
As a result, a straightforward approach is to treat all channel entries in these channel matrices as independent elements and solve linear functions formulated by stacking the received signals from multiple time instants. 
Since the number of elements in $\mathbf{J}_{1,k}$'s, $\mathbf{J}_{2,k}$'s, and $\mathbf{J}_{1,2,k}$'s is as large as $KL(M_1^2 + M_2^2 + M_1^2M_2^2)$, the minimum training duration required by this method is $K(M_1^2 + M_2^2 + M_1^2M_2^2)$, scaling prohibitively with the product of the number of users and the squared dimensions of the two BD-RISs. 
In our recent work \cite{ref:wangrui}, we revealed that in a single-BD-RIS-aided system, the entries in user channels are correlated, and such correlation can be adopted to significantly reduce the channel estimation overhead. In this work, we aim to generalize such a result to a double-BD-RIS-aided system, 
where the presence of two BD-RISs gives rise to more intricate channel correlations.

To achieve this goal, we first reveal that the high-dimensional cascaded channels, i.e., $\mathbf{J}_{1,k}$'s, $\mathbf{J}_{2,k}$'s, and $\mathbf{J}_{1,2,k}$'s, can be fully characterized by five matrices of much lower dimensions. Specifically, define two non-zero coefficients for BD-RISs 1 and 2 as\footnote{
In \cite{ref:wangrui}, the scalar $c_i$ is set to be $r_{i,1,1}$, and thus the cascaded channels are recovered by estimating one typical user's channel and a set of coefficients associated with the remaining users. A key limitation of this approach is that it requires other users to remain silent during the estimation of the typical user's channel, which is ineffective especially in the scenario with low SNR. To address this issue, as shown in \eqref{eq:barQ1} and \eqref{eq:Q1kn_2_c1}, we propose to treat the linear combination of the channels from all users to an element of the BD-RIS and then to the BS as the reference channel $\bar{\mathbf{Q}}_{i}$. 
This formulation allows all users to transmit pilots simultaneously, thereby improving the received SNR 
in practical noisy environments, as will be introduced in Section \ref{Sec:CE_AWGN}.}
\begin{equation} \label{eq:Q1kn_2_c1}
  c_i = {\sum}_{k = 1}^{K} r_{i,k,1}, \;\; \forall  i \in \{1,2\}. 
\end{equation} 
Then, for single-reflection links, 
the sub-blocks $\mathbf{Q}_{i,k,m_i}$'s of $\mathbf{J}_{i,k}$ in \eqref{eq:J1k} satisfy that 
\begin{equation} \label{eq:Q1kn_2}
  \mathbf{Q}_{i,k,m_i} = \bar{r}_{i,k,m_i} \bar{\mathbf{Q}}_{i}, \;\; \forall k,m_i, i\in \{1,2\},
\end{equation}
where
\begin{equation} \label{eq:barr1kn}
  \bar{r}_{i,k,m_i} = \tfrac{r_{i,k,m_i}}{c_i} , \;\; \forall  k,m_i, i\in \{1,2\},
\end{equation}
\begin{equation} \label{eq:barQ1}
  \bar{\mathbf{Q}}_{i} = c_i \mathbf{G}_{i} , \;\; \forall  i\in \{1,2\}. 
\end{equation} 
For convenience, define $\bar{\bm{r}}_{i} = [\bar{r}_{i,1,2}, \ldots, \bar{r}_{i,1,M_i}, \ldots, \bar{r}_{i,K,1},$ $\ldots, \bar{r}_{i,K,M_i}]^T, \forall i\in\{1,2\}$. Therefore, it is sufficient to estimate $\bar{\bm{r}}_{i}$'s and $\bar{\mathbf{Q}}_{i}$'s to recover the single-reflection channels $\mathbf{J}_{i,k}$'s based on \eqref{eq:J1k}, \eqref{eq:Q1kn_2}, and 
\begin{equation} \label{eq:barr1kn11}
  \bar{r}_{i,1,1} = 1 - {\sum}_{k \neq 1} \bar{r}_{i,k,1} , \;\; \forall  i \in \{1,2\}. 
\end{equation} 
On the other hand, according to \eqref{eq:J12k}, the sub-blocks $\mathbf{Q}_{1,2,k,m_2,m_1,m'_1}$'s of the double-reflection channel $\mathbf{J}_{1,2,k}$ in \eqref{eq:J12k} satisfy that 
\begin{equation} \label{eq:Q12kmnn_2} 
  \mathbf{Q}_{1,2,k,m_2,m_1,m'_1} \!=  \bar{b}_{m_2,m_1} \bar{r}_{1,k,m'_1} \bar{\mathbf{Q}}_{2}, \;  \forall k,m_1,m'_1,m_2,  
\end{equation}
where
\begin{equation} 
  \bar{b}_{m_2,m_1} = \tfrac{b_{m_2,m_1} c_1}{c_2} , \;\; \forall m_1,m_2.  \label{eq:Q12kmnn_2_bmn}
\end{equation}
Denote $\bar{\mathbf{B}} = \frac{c_1}{c_2} \mathbf{B}$, whose $(m_2,m_1)$-th element is $\bar{b}_{m_2,m_1}$ in \eqref{eq:Q12kmnn_2_bmn}. 
Then, we can reconstruct the double-reflection cascaded channels $\mathbf{J}_{1,2,k}$'s by estimating $\bar{\mathbf{Q}}_{2}, \bar{\mathbf{B}}$, and $\bar{\bm{r}}_{1}$ and then substituting the estimates into \eqref{eq:J12k}, \eqref{eq:barr1kn11}, and \eqref{eq:Q12kmnn_2}. 
To summarize, to recover $\mathbf{J}_{1,k}$'s, $\mathbf{J}_{2,k}$'s, and $\mathbf{J}_{1,2,k}$'s, it is sufficient to recover five matrices:  $\bar{\mathbf{Q}}_{1}, \bar{\mathbf{Q}}_{2}, \bar{\mathbf{B}}, \bar{\bm{r}}_{1}$, and $\bar{\bm{r}}_{2}$. 
In this way, the number of unknowns is reduced from $KL(M_1^2 + M_2^2 + M_1^2M_2^2)$ in $\mathbf{J}_{1,k}$'s, $\mathbf{J}_{2,k}$'s, and $\mathbf{J}_{1,2,k}$'s to $(L+K)(M_1 + M_2) + M_1M_2 - 2$ in $\bar{\mathbf{Q}}_{1}, \bar{\mathbf{Q}}_{2}, \bar{\mathbf{B}}, \bar{\bm{r}}_{1}$, and $\bar{\bm{r}}_{2}$. 
This indicates that 
it is possible to estimate the cascaded channels with much lower overhead compared to treating all of their entries as independent variables and solving linear functions formulated by stacking the received signals from multiple time instants.

The substantial reduction in the number of elements to be estimated stems from three observations. 
First, similar to \cite{ref:wangrui}, for the single-reflection links, BD-RIS $i \in \{1,2\}$ reflects incoming signals towards the BS through a shared BD-RIS $i$ - BS link, and thus the matrix $\mathbf{Q}_{i,k,m_i}$ is a scaled version of the reference channel, as shown in \eqref{eq:Q1kn_2}. 
Here, the reference channel is selected as the linear combination of the channels from all users to an element of the BD-RIS and then to the BS, i.e. $\bar{\mathbf{Q}}_{i}$, instead of one typical user's channel adopted in \cite{ref:wangrui}, as introduced before. 
Second, in the double-reflection link, BD-RIS $1$ reflects the incoming signals (from any user $k \in [K]$ to any element $m'_1 \in [M_1]$ of BD-RIS 1) through the shared channel towards BD-RIS $2$, and BD-RIS $2$ reflects the incoming signals (from any element $m_1 \in [M_1]$ of BD-RIS 1 to any element $m_2 \in [M_2]$ of BD-RIS 2) through the shared channel towards the BS. Thus, as shown in \eqref{eq:Q12kmnn_2}, $\mathbf{Q}_{1,2,k,m_2,m_1,m'_1}$ is a scaled version of a reference matrix, with the scaling coefficient dependent on the indices $k,m_1,m'_1,$ and $m_2$. 
Third, the channel corresponding to the single-reflection link (via BD-RIS 1 or BD-RIS 2 only) and that corresponding to the double-reflection link (via both BD-RIS 1 and BD-RIS 2) are highly correlated. Specifically, the single-reflection link via BD-RIS 1 and the double-reflection link share the same user $k$ - BD-RIS 1 channel, and thus $\bar{\bm{r}}_{1}$ is incorporated in both $\mathbf{J}_{1,k}$ and $\mathbf{J}_{1,2,k}$, $\forall k\in[K]$, as shown in \eqref{eq:Q1kn_2} and \eqref{eq:Q12kmnn_2}. Likewise, the single-reflection channel via BD-RIS 2 and the double-reflection channel share the common BD-RIS 2 - BS link, and thus $\bar{\mathbf{Q}}_2$ is incorporated in both $\mathbf{J}_{2,k}$ and $\mathbf{J}_{1,2,k}$, $\forall k\in[K]$. 
Building on above channel correlation properties, we can characterize the high-dimensional cascaded channels by five low-dimensional matrices as mentioned before, thereby dramatically reducing the number of elements to estimate.


However, how to estimate $\bar{\mathbf{Q}}_{1}, \bar{\mathbf{Q}}_{2}, \bar{\mathbf{B}}, \bar{\bm{r}}_{1}$, and $\bar{\bm{r}}_{2}$ remains a significant challenge. 
To illustrate this point, we follow the above formulation and re-express the received signal $\mathbf{y}_t$ 
as 
\begin{align}  
  \mathbf{y}_t  
  = & \sum_{k=1}^{K}  \sqrt{p} x_{k,t}  \left( \sum_{m'_1= 1}^{M_1} \!\bar{r}_{1,k,m'_1} \bar{\mathbf{Q}}_{1} [\bm{\Phi}_{1,t}]_{:,m'_1} \right. \notag\\
  & +  \sum_{m_2 =1}^{M_2}  \bar{r}_{2,k,m_2} \bar{\mathbf{Q}}_{2} [\bm{\Phi}_{2,t}]_{:,m_2}  
  + \sum_{m'_1 =1}^{M_1}  \sum_{m_2 =1}^{M_2} \sum_{m_1 =1}^{M_1}   \bar{b}_{m_2,m_1}    \notag\\
  &  \times \bar{r}_{1,k,m'_1} \bar{\mathbf{Q}}_{2}  [\bm{\Phi}_{1, t}]_{m_1,m'_1} [\bm{\Phi}_{2,t}]_{:,m_2}  
   \Big)  + \mathbf{z}_t. 
  \label{eq:yt_2} 
\end{align}  
As we can see, the key challenge lies in that the received signal is no longer linear with $\bar{\mathbf{Q}}_{1}, \bar{\mathbf{Q}}_{2}, \bar{\mathbf{B}}, \bar{\bm{r}}_{1}$, and $\bar{\bm{r}}_{2}$ due to the product terms $\bar{r}_{1,k,m'_1} \bar{\mathbf{Q}}_{1}$, $\bar{r}_{2,k,m_2} \bar{\mathbf{Q}}_{2}$, and $\bar{b}_{m_2,m_1} \bar{r}_{1,k,m'_1} \bar{\mathbf{Q}}_{2}$. 
It should be noted that the coupling between the unknown parameters in the double-BD-RIS channel is more complicated compared to the single-BD-RIS channel considered in \cite{ref:wangrui}. 
Specifically, in the system with only BD-RIS 1, the nonlinearity stems from the coupling term $\bar{r}_{1,k,m'_1} \bar{\mathbf{Q}}_{1}$. However, in the double-BD-RIS case, $\bar{r}_{1,k,m'_1}$ is also coupled with $\bar{b}_{m_2,m_1}$ and $\bar{\mathbf{Q}}_{2}$. 
If there is only BD-RIS 2, the nonlinearity only arises from the coupling term $\bar{r}_{2,k,m_2} \bar{\mathbf{Q}}_{2}$. In contrast, as shown in \eqref{eq:yt_2}, $\bar{\mathbf{Q}}_{2}$ is also coupled with $\bar{b}_{m_2,m_1}$ and $\bar{r}_{1,k,m'_1}$. 
Moreover, compared to the double-diagonal-RIS system considered in \cite{ref:doubleRIS_ZhengBY_TCOM}, the scattering matrices in the double-BD-RIS system should satisfy the unitary constraint given in \eqref{eq:Phi}. This requirement introduces additional coupling associated with off-diagonal elements of BD-RIS scattering matrices. The enhanced coupling complicates the theoretical analysis and algorithm design in double-BD-RIS networks. 
To tackle this issue, in the rest of this paper, we will first consider the ideal noiseless case and develop an advanced channel estimation framework by fully exploiting the aforementioned channel properties. 
A comprehensive analysis of the required training overhead will be provided. 
Then, the proposed channel estimation framework will be extended to the practical case with noise at the BS.

\section{Channel Estimation in the Noiseless Case} \label{Sec:CE_noAWGN}


In this section, we design a low-overhead channel estimation framework for the ideal noiseless case and establish its theoretical limits.
The estimation process is divided into five phases to sequentially estimate $\bar{\mathbf{Q}}_{2}, \bar{\bm{r}}_{2}, \bar{\mathbf{Q}}_{1}, \bar{\mathbf{B}},$ and $\bar{\bm{r}}_{1}$. 
Let $\tau_d$ be the time duration of phase $d$, and $\tau'_d = \sum_{d'=1}^{d} \tau_{d'}$ be the cumulative time. 
To estimate the $d$-th matrix, we construct linear functions with respect to it by treating the previously estimated $d-1$ matrices as known and eliminating the effect of the remaining $5-d$ matrices. 
This decoupling effect is achieved by jointly designing the scattering matrices of two BD-RISs and the pilots of $K$ users over time, as will be introduced in the following.

\subsection{Phase I: Estimation of $\bar{\mathbf{Q}}_{2}$} \label{Sec:IV-I}

As shown in \eqref{eq:yt_2}, the estimation of $\bar{\mathbf{Q}}_{2}$ is challenging due to its coupling with $\bar{\mathbf{r}}_2$, $\bar{\mathbf{B}}$, and $\bar{\mathbf{r}}_1$, as well as the interference resulting from the single-reflection link via BD-RIS 1. 
To address these issues, we design the pilot signals and BD-RIS scattering matrices as follows. 
Specifically, the total duration $\tau_1$ used to estimate $\bar{\mathbf{Q}}_{2}$ is divided into four parts of the same length $\tau_{1,1}$, i.e., $\tau_1 = 4 \tau_{1,1}$. 
At each time instant $t = \tau \in [\tau_{1,1}]$ of the first part, all $K$ users are assigned a common, arbitrary scalar with unit power as the pilot signal, i.e., 
\begin{equation} \label{eq:I1_x}
  x_{k,\tau} = x_{\tau}, \;\; \forall k\in[K], \tau \in [\tau_{1,1}], 
\end{equation}
satisfying $|x_{\tau}|=1, \forall \tau \in [\tau_{1,1}]$. 
Let $\mathbf{D}\in \mathbb{C}^{M_1\times M_1}$ be an arbitrary unitary matrix. The scattering matrix $\bm{\Phi}_{1,\tau}$ of BD-RIS 1 in the first part is set as
\begin{equation} \label{eq:I1_Psi}
  \bm{\Phi}_{1,\tau} = \mathbf{D}, \;\; \forall \tau \in [\tau_{1,1}]. 
\end{equation}
Let $\mathbf{P} = [\mathbf{p}_1, \ldots, \mathbf{p}_{M_2}] \in \mathbb{C}^{M_2\times M_2}$ be an arbitrary unitary matrix. Then, the $m_2$-th column of the scattering matrix $\bm{\Phi}_{2,\tau}$ of BD-RIS 2 in the first part is set as 
\begin{equation} \label{eq:I1_Phi}
  [\bm{\Phi}_{2,\tau}]_{:,m_2} = \mathbf{p}_{((m_2+{\tau}-2) \operatorname{mod} M_2)+1}, \; \forall m_2\in[M_2], \tau \in [\tau_{1,1}]. 
\end{equation}

In the second part, i.e., at each time instant $t = \tau_{1,1} + \tau \in [\tau_{1,1}+1, 2\tau_{1,1}]$, the pilot signal $x_{k,\tau_{1,1}+\tau}$ of user $k$ and the scattering matrices $\bm{\Phi}_{1,\tau_{1,1}+\tau}$ and $\bm{\Phi}_{2,\tau_{1,1}+\tau}$ are set as
\begin{equation} \label{eq:I2_x}
  x_{k,\tau_{1,1}+\tau} = x_{k,\tau}, \;\; \forall k\in[K], \tau \in [\tau_{1,1}], 
\end{equation} 
\begin{equation}  \label{eq:I2_Psi}
  \bm{\Phi}_{1,\tau_{1,1}+\tau} = - \bm{\Phi}_{1,\tau}, \;\; \forall \tau \in [\tau_{1,1}],  
\end{equation}  
\begin{equation}  \label{eq:I2_Phi}
  \bm{\Phi}_{2,\tau_{1,1}+\tau} = \bm{\Phi}_{2,\tau}, \;\; \forall \tau \in [\tau_{1,1}].  
\end{equation} 

In the third part, i.e., at each time instant $t = 2\tau_{1,1} + \tau \in [2\tau_{1,1}+1, 3\tau_{1,1}]$, we set pilot signals and scattering matrices~as 
\begin{equation} \label{eq:I3_x}
  x_{k,2\tau_{1,1}+\tau} = x_{k,\tau} , \;\; \forall k\in[K], \tau \in [\tau_{1,1}], 
\end{equation}
\begin{equation}  \label{eq:I3_Psi}
  \bm{\Phi}_{1,2\tau_{1,1}+\tau} = \bm{\Phi}_{1,\tau} , \;\; \forall \tau \in [\tau_{1,1}], 
\end{equation}
\begin{equation} \label{eq:I3_Phi}
  [\bm{\Phi}_{2,2\tau_{1,1}+\tau}]_{:,m_2} \!=\! \left\{ \!\! \begin{array}{ll} 
  e^{\jmath \theta} [\bm{\Phi}_{2,\tau}]_{:,1}, \!\!\!& m_2 = 1,  \\
  {[\bm{\Phi}_{2,\tau}]_{:,m_2}} , & m_2 \in [2, M_2], 
  \end{array} \right. \!\!\!\; \forall \tau \in [\tau_{1,1}], 
\end{equation}  
where $\theta\in(0, 2\pi)$ is an arbitrary phase shift. 

In the fourth part, i.e., at each time instant $t = 3\tau_{1,1} + \tau \in [3\tau_{1,1}+1, 4\tau_{1,1}]$, we set pilot signals and scattering matrices~as
\begin{equation} \label{eq:I4_x}
  x_{k,3\tau_{1,1}+\tau} = x_{k,\tau} , \;\; \forall k\in[K], \tau \in [\tau_{1,1}], 
\end{equation} 
\begin{equation}  \label{eq:I4_Psi}
  \bm{\Phi}_{1,3\tau_{1,1}+\tau} = -\bm{\Phi}_{1,\tau} , \;\; \forall \tau \in [\tau_{1,1}], 
\end{equation} 
\begin{equation} \label{eq:I4_Phi}
  \bm{\Phi}_{2,3\tau_{1,1}+\tau} = \bm{\Phi}_{2,2\tau_{1,1}+\tau},  \; \forall \tau \in [\tau_{1,1}].
\end{equation}    

Substituting the designed pilot signals and scattering matrices into \eqref{eq:yt_2} and removing the noise, we can obtain the received signals in the four parts, which are denoted as $\mathbf{y}_{\tau}, \mathbf{y}_{\tau_{1,1}+\tau}, \mathbf{y}_{2\tau_{1,1}+\tau},$ and $\mathbf{y}_{3\tau_{1,1}+\tau}, \forall \tau\in[\tau_{1,1}]$, respectively. Then, we can construct the effective received signal given by 
\begin{align}  
  \bar{\mathbf{y}}_{1,\tau} & = \tfrac{ \mathbf{y}_{\tau} + \mathbf{y}_{\tau_{1,1}+\tau} }{2} - \tfrac{\mathbf{y}_{2\tau_{1,1}+\tau} + \mathbf{y}_{3\tau_{1,1}+\tau}}{2} \notag \\
  & = c_{\theta} x_{\tau} \bar{\mathbf{Q}}_{2}  [\bm{\Phi}_{2,\tau}]_{:,1}  ,  \;\;  \forall  \tau \in [ \tau_{1,1}] , \label{eq:bary1t}
\end{align}
where 
\begin{equation} \label{eq:c_theta}
  c_{\theta} = \sqrt{p} (1-e^{\jmath \theta}). 
\end{equation} 
In the case of $\tau_{1,1}=M_2$, i.e. $\tau_{1}=4M_2$, we can perfectly estimate $\bar{\mathbf{Q}}_{2}$ based on the overall effective received signal 
\begin{equation} \label{eq:bary1}
  \bar{\mathbf{Y}}_1 = [ \bar{\mathbf{y}}_{1,1} , \ldots , \bar{\mathbf{y}}_{1,M_2} ] = c_{\theta} \bar{\mathbf{Q}}_{2}  \mathbf{A}_1 , 
\end{equation}
where the matrix $\mathbf{A}_1 = [ x_{1} [\bm{\Phi}_{2,1}]_{:,1} , \ldots , x_{M_2} [\bm{\Phi}_{2,M_2}]_{:,1} ]$ 
has rank $M_2$. 
Then, $\bar{\mathbf{Q}}_{2}$ can be perfectly estimated as
\begin{equation} \label{eq:Q2_perfect_est}
  \bar{\mathbf{Q}}_{2} = c_{\theta}^{-1} \bar{\mathbf{Y}}_1  \mathbf{A}_1^H (\mathbf{A}_1 \mathbf{A}_1^H)^{-1} . 
\end{equation}

\subsection{Phase II: Estimation of $\bar{\bm{r}}_{2}$} \label{Sec:IV-II}

%
%

Denote the number of time instants used to estimate $\bar{\bm{r}}_{2}$ as $\tau_2$, which is divided into two parts with the same length $\tau_{2,1}$, i.e. $\tau_2 = 2 \tau_{2,1}$. 
At each time instant $t = \tau_1 + \tau \in [\tau_1 + 1, \tau_1 +  \tau_{2,1}]$ of the first part, we design the pilot signal $x_{k,\tau_{1}+\tau}$ of user $k \in [K]$, the scattering matrix $\bm{\Phi}_{1,\tau_{1}+\tau}$ of BD-RIS 1, and the scattering matrix $\bm{\Phi}_{2,\tau_{1}+\tau}$ of BD-RIS 2 as follows. 
Specifically, the scattering matrix $\bm{\Phi}_{1,\tau_{1}+t}$ is set as  
\begin{equation} \label{eq:psi_r2}
  \bm{\Phi}_{1,\tau_{1}+\tau} = \mathbf{D}, \;\; \forall \tau \in [\tau_{2,1}]. 
\end{equation}
where $\mathbf{D}\in \mathbb{C}^{M_1\times M_1}$ denotes an arbitrary unitary matrix as aforementioned. 
The scattering matrix $\bm{\Phi}_{2,\tau_{1}+\tau}$ is designed based on the estimate of $\bar{\mathbf{Q}}_{2}$ in \eqref{eq:Q2_perfect_est}. 
Denote the rank of $\bar{\mathbf{Q}}_{2}$ as $q_2$. The singular value decomposition (SVD) of $\bar{\mathbf{Q}}_{2}$ is expressed as $\bar{\mathbf{Q}}_{2} = \mathbf{U}_{\bar{\mathbf{Q}}_2} \bm{\Sigma}_{\bar{\mathbf{Q}}_2} \mathbf{V}_{\bar{\mathbf{Q}}_2}^H$, where $\mathbf{U}_{\bar{\mathbf{Q}}_2} \in \mathbb{C}^{L\times L}$ and $\mathbf{V}_{\bar{\mathbf{Q}}_2} \in \mathbb{C}^{M_2\times M_2}$ are unitary matrices, 
and $ \bm{\Sigma}_{\bar{\mathbf{Q}}_{2 }} =  \operatorname{blkdiag} \{ \bm{\Lambda}_{\bar{\mathbf{Q}}_{2 }} , \bm{0}_{L-q_2,M_2-q_2}\}$ is an $L\times M_2$ rectangular diagonal matrix 
with $\bm{\Lambda}_{\bar{\mathbf{Q}}_{2 }} = \operatorname{diag}\{\sigma_{\bar{\mathbf{Q}}_{2 },1} , \ldots, \sigma_{\bar{\mathbf{Q}}_{2 },q_2} \}$ containing $q_2$ nonzero singular values. 
Then, we design $\bm{\Phi}_{2,\tau_1 + \tau}$ as 
\begin{equation} \label{eq:phi_r2}
  \bm{\Phi}_{2,\tau_1 + \tau} = \mathbf{V}_{\mathbf{Q}_2} \mathbf{P}_{\tau_1 + \tau}, \;\; \forall \tau \in [\tau_{2,1}], 
\end{equation}
where $\mathbf{P}_{\tau_1 + \tau}$ is an $M_2\times M_2$ unitary matrix.

Next, we aim to design the unitary matrix ${\mathbf{P}}_{\tau_1 + \tau}$ and the pilot signal $\mathbf{x}_{\tau_1 + \tau} = [x_{1,\tau_1 + \tau}, \ldots, x_{K,\tau_1 + \tau}]^T$ at each time instant $t=\tau_1 + \tau \in [\tau_1 + 1, \tau_1 +  \tau_{2,1}]$, i.e., for each $\tau\in[\tau_{2,1}]$, following the approach proposed in \cite{ref:wangrui}. 
Specifically, define $K_0$ as the smallest integer such that $\tau_0 = \frac{M_2K_0}{q_2}$ is an integer. Denote $\eta = \lfloor \frac{K}{K_0} \rfloor$. 
For each $\tau\in[\tau_{2,1}]$, we divide $K$ elements in $\mathbf{x}_{\tau_1 + \tau}$ into $\eta+1$ groups, with the size of each group given~by 
\begin{equation} 
  K_n = \left\{ \begin{array}{ll}
  K_0, & n \in [\eta], \\
  K - \eta K_0 , & n = \eta + 1. 
  \end{array} \right.
\end{equation}
In the case of $K \operatorname{mod} K_0 = 0$, the size of the last group is equal to $0$. 
Denote the set of user indexes in each group as 
\begin{equation} 
  \mathcal{K}_n = \left\{ \begin{array}{ll}
  \{ (n-1) K_0 + 1, \ldots, n K_0 \}, & n \in [\eta] , \\
  \{ \eta K_0 + 1, \ldots, K \} , & n = \eta + 1.  
  \end{array} \right.
\end{equation}
Correspondingly, we divide $\tau_{2,1}$ time instants into $\eta+1$ non-overlapping groups, with the size of each group given by 
\begin{equation} \label{eq:Ti_r2}
  T_n = \left\{ \begin{array}{ll}
  \tau_0, & n \in [\eta], \\
  \tau_{2,1} - \eta \tau_0 , & n = \eta + 1. 
  \end{array} \right.
\end{equation}
Denote the set of time instants in each of the $\eta+1$ groups as 
\begin{equation} \label{eq:Tiset_r2}
  \mathcal{T}_n = \left\{ \! \begin{array}{ll} 
  \tau_1 + \{ (n-1) \tau_0 + 1, \ldots, n \tau_0 \}, & n \in [\eta] , \\
  \tau_1 + \{ \eta \tau_0 + 1, \ldots, \tau_{2,1} \} , & n = \eta + 1.  
  \end{array} \right.
\end{equation}
For each $n \in [\eta+1]$ and $j \in [K_n]$, we define the set $\mathcal{R}_{n,j}$ as
%
\begin{equation} \label{eq:Rij_r2}
  \mathcal{R}_{n,j} = \left\{  \begin{array}{ll} 
  \left\{ 1 , \ldots, \lceil \frac{M_2}{q_2} \rceil \right\}, &  j = 1, \\
  \left\{ \lceil \frac{(j-1)M_2}{q_2} \rceil , \ldots, \lceil \frac{jM_2}{q_2} \rceil \right\} , &  j \in [2,K_n]. 
  \end{array} \right.  
\end{equation}
Then, the pilot signal is designed as
\begin{align}
  x_{ \mathcal{K}_{n}(j) , \mathcal{T}_{n'}(r) } & = \left\{ \begin{array}{ll}
  c_{\mathcal{T}_n(r) }, & \text{if} \;\; n = n' \; {\rm{and}} \; r \in \mathcal{R}_{n,j}, \\
  0 , & \text{otherwise}, 
  \end{array} \right.   \notag\\
  & \!\!\!\!\!\!\!\!\!\!\!\! \forall n\in [\eta+1], n' \in [\eta+1],  j\in[K_n], r\in [T_{n'}],  \label{eq:x_r2}
\end{align}
where $c_{\mathcal{T}_n(r) }$ is an arbitrary unit-power scalar. 
Recalling that $\mathbf{P} \in \mathbb{C}^{M_2\times M_2}$ is an arbitrary unitary matrix. Denote the $m_2$-th row of $\mathbf{P}$ as $\bar{\mathbf{p}}_{m_2}^T$. Then, the matrix $\mathbf{P}_{\mathcal{T}_n(r)}$ is designed as 
\begin{align} 
  \mathbf{P}_{\mathcal{T}_n(r)}\! = & \left[ \bar{\mathbf{p}}_{ ((r - 1) q_2 \; \operatorname{mod} \; {M_2}) +1} , \ldots, \bar{\mathbf{p}}_{M_2}, \bar{\mathbf{p}}_{1} , \ldots,  \right.\notag\\
  & \; \left. \bar{\mathbf{p}}_{ ((r - 1) q_2 \; \operatorname{mod} \; {M_2}) } \right]^T\!, 
  \; \forall n\!\in\![\eta+1], r \!\in\! [T_n]. \label{eq:P_r2}
\end{align}  
In the second part, i.e., at each time instant $t = \tau_1 + \tau_{2,1} + \tau \in [\tau_1\! +\! \tau_{2,1} \!+\! 1, \tau_1\! +\!  2\tau_{2,1}]$, we set pilots and scattering matrices~as
\begin{equation} \label{eq:II2_x}
  \mathbf{x}_{\tau_1 + \tau_{2,1} + \tau} = \mathbf{x}_{\tau_1 + \tau} , \;\; \forall \tau \in [\tau_{2,1}],   
\end{equation} 
\begin{equation}  \label{eq:II2_Psi}
  \bm{\Phi}_{1,\tau_1 + \tau_{2,1} + \tau} = -\bm{\Phi}_{1,\tau_1 + \tau} , \;\; \forall \tau \in [\tau_{2,1}],   
\end{equation}  
\begin{equation} \label{eq:II2_Phi}
  \bm{\Phi}_{2,\tau_1 + \tau_{2,1} + \tau} = \bm{\Phi}_{2,\tau_1 + \tau}, \;\; \forall \tau \in [\tau_{2,1}]. 
\end{equation}

Substituting the designed pilot signals and scattering matrices into \eqref{eq:yt_2} and removing the noise, we can obtain the received signals in the two parts, denoted as $\mathbf{y}_{\tau_1 + \tau}$ and $\mathbf{y}_{\tau_1 + \tau_{2,1} + \tau}$, $\forall \tau\in[ \tau_{2,1}]$, respectively. Then, we can construct the effective received signal as   
\begin{align}  
  & \bar{\mathbf{y}}_{2,\tau} 
  = \tfrac{ \mathbf{y}_{\tau_1 + \tau} + \mathbf{y}_{\tau_1 + \tau_{2,1} + \tau} }{2} \notag \\
  & = \!{\sum}_{k=1}^{K} {\sum}_{m_2=1}^{M_2} \sqrt{p}    x_{k,\tau_1 + \tau}  \bar{r}_{2,k,m_2} \bar{\mathbf{Q}}_{2} [\bm{\Phi}_{2,\tau_1 + \tau}]_{:,m_2} \notag \\
  & = \!\! \sqrt{p}\! \left(  \mathbf{x}_{\tau_1 + \tau}^T \!  \otimes \! \left( \bar{\mathbf{Q}}_{2} \bm{\Phi}_{2, \tau_1 + \tau} \right) \right)\! [ \bar{r}_{2,1,1}, \bar{\bm{r}}_{2}^T]^T \! ,   \;   \forall   \tau\! \in \![ \tau_{2,1}] . 
\end{align} 
The overall effective received signal is given by 
\begin{equation} 
  \bar{\mathbf{y}}_2 = [ \bar{\mathbf{y}}_{2,1}^T , \ldots , \bar{\mathbf{y}}_{2,\tau_{2,1}}^T ]^T
  = \sqrt{p} \mathbf{A}_2  [ \bar{r}_{2,1,1}, \bar{\bm{r}}_{2}^T]^T  , 
\end{equation}
where
\begin{equation} \label{eq:A2}
  \mathbf{A}_2 = \begin{bmatrix}
  \mathbf{x}_{\tau_1 + 1}^T \otimes \left( \bar{\mathbf{Q}}_{2} \bm{\Phi}_{2,\tau_1 + 1} \right) \\
  \vdots \\
  \mathbf{x}_{\tau_1 +\tau_{2,1}}^T \otimes \left( \bar{\mathbf{Q}}_{2} \bm{\Phi}_{2,\tau_1 + \tau_{2,1}} \right)
  \end{bmatrix} .
\end{equation}
As proved in \cite[Theorem 2]{ref:wangrui}, following the design criterion in \eqref{eq:phi_r2}, \eqref{eq:x_r2}, and \eqref{eq:P_r2} with $\tau_{2,1} = \lceil \frac{KM_2}{q_2} \rceil$, the matrix $\mathbf{A}_2$ has full column rank, i.e., $\operatorname{rank}(\mathbf{A}_2) = KM_2$. 
Then, $\bar{\bm{r}}_{2}$ can be perfectly estimated as
\begin{equation} 
  \bar{\bm{r}}_{2} = ( \sqrt{p} )^{-1} [ (  \mathbf{A}_2^H \mathbf{A}_2 )^{-1} ]_{2:KM_2,:} \mathbf{A}_2^H \bar{\mathbf{y}}_2  .  
\end{equation}

\subsection{Phase III: Estimation of $\bar{\mathbf{Q}}_{1}$} \label{Sec:IV-III}

To avoid the coupling with $\bar{\mathbf{r}}_1$ and the interference from the single-reflection of BD-RIS 2, 
we divide the time duration $\tau_3$ in this phase into four parts of equal length $\tau_{3,1}$, i.e. $\tau_3 = 4 \tau_{3,1}$, and design the pilot signals and scattering matrices in the four parts as 
\begin{align} 
  & x_{k,\tau'_2 + \tau} \!\!=\! x_{k,\tau'_2 + \tau_{3,1}+\tau} \!\!=\! x_{k,\tau'_2 + 2\tau_{3,1}+\tau} \!\!=\! x_{k,\tau'_2 + 3\tau_{3,1}+\tau} \!\!=\! x_{\tau'_2 + \tau},  \notag \\
  & \;\;\;\;\;\;\; \;\;\;\;\;\;\; \;\;\;\;\;\;\;  \forall k\in[K], \tau\in[\tau_{3,1}], \label{eq:Psi_Q1}
\end{align} 
\begin{align} 
  & [\bm{\Phi}_{1,\! \tau'_2 + \tau}]_{:,m_1} \!=\! [\bm{\Phi}_{1,\tau'_2 + \tau_{3,1}+\tau}]_{:,m_1}  \notag\\
  & \!=\! [\mathbf{D}]_{ :,{((m_1+\tau-2) \operatorname{mod} M_1)+1} }, 
  \;\; \forall m_1 \in [M_1], \tau \in [\tau_{3,1}], 
\end{align}  
\begin{align}  
  & [\bm{\Phi}_{1,\tau'_2 + 2\tau_{3,1}+\tau}]_{:,m_1} \! = 
  [\bm{\Phi}_{1,\tau'_2 + 3\tau_{3,1}+\tau}]_{:,m_1} \notag \\
  & = \left\{ \begin{array}{ll} 
  e^{\jmath \theta} [\bm{\Phi}_{1,\tau'_2 + \tau}]_{:,1}, &\!\! m_1 = 1,  \\
  {[\bm{\Phi}_{1,\tau'_2 + \tau}]_{:,m_1}} , & \!\!m_1 \in [2, M_1], 
  \end{array} \right.  \forall \tau\in[\tau_{3,1}], 
\end{align} 
\begin{align} 
  & \bm{\Phi}_{2,\tau'_2 + \tau} \!\!=\!\! - \bm{\Phi}_{2,\tau'_2 + \tau_{3,1}+\tau} \!\!= \!\!\bm{\Phi}_{2,\tau'_2 + 2\tau_{3,1}+\tau} \!\!=\!\! -\bm{\Phi}_{2,\tau'_2 + 3\tau_{3,1}+\tau} \!\!= \!\mathbf{P}, \notag\\
  & \;\;\;\;\;\;\; \;\;\;\;\;\;\; \;\;\;\;\;\;\;  \forall \tau\in[\tau_{3,1}], 
\end{align}
where $x_{\tau'_2 +\tau}$ is an arbitrary unit-power scalar, $\mathbf{D}$ and $\mathbf{P}$ are arbitrary $M_1\times M_1$ and $M_2\times M_2$ unitary matrices, respectively, and $\theta\in(0, 2\pi)$ is an arbitrary phase shift, as defined before. 

Based on the specific settings over time instants $t \in [\tau'_2 + 1, \tau'_3]$, we can obtain the received signals in the four parts, which are denoted as $\mathbf{y}_{\tau'_2 + \tau}, \mathbf{y}_{\tau'_2 + \tau_{3,1}+\tau}, \mathbf{y}_{\tau'_2 + 2\tau_{3,1}+\tau},$ and $\mathbf{y}_{\tau'_2 +3\tau_{3,1}+\tau}, \forall \tau\in[\tau_{3,1}]$, respectively. Then, similar to in Phase I, we can construct the effective received signal in Phase III, given by
\begin{subequations}
\begin{align}  
  \bar{\mathbf{y}}_{3,\tau} 
  & = \tfrac{  \mathbf{y}_{\tau'_2 + \tau} + \mathbf{y}_{\tau'_2 + \tau_{3,1}+\tau} }{2} - \tfrac{ \mathbf{y}_{\tau'_2 + 2\tau_{3,1}+\tau} + \mathbf{y}_{\tau'_2 + 3\tau_{3,1}+\tau} }{2}  \\
  & = c_{\theta} x_{\tau'_2 +\tau} \bar{\mathbf{Q}}_{1}  [\bm{\Phi}_{1,\tau'_2 +\tau}]_{:,1}  ,  \;\; \forall  \tau \in [ \tau_{3,1}] . 
\end{align}
\end{subequations}
In the case of $\tau_{3,1}=M_1$, i.e. $\tau_{3}=4M_1$, we can perfectly estimate $\bar{\mathbf{Q}}_{1}$ based on the overall effective received signal given by
\begin{equation} \label{eq:yt_noAWGN_singleUE_Q111_all}
  \bar{\mathbf{Y}}_3 = [ \bar{\mathbf{y}}_{3,1} , \ldots , \bar{\mathbf{y}}_{3,M_1} ] = c_{\theta} \bar{\mathbf{Q}}_{1} \mathbf{A}_3 , 
\end{equation}
where $\mathbf{A}_3 = [ x_{\tau'_2 +1} [\bm{\Phi}_{1,\tau'_2 + 1}]_{:,1} , \ldots , x_{\tau'_2 +M_1} [\bm{\Phi}_{1,\tau'_2 + M_1}]_{:,1} ] $ has rank $M_1$. 
Then, $\bar{\mathbf{Q}}_{1}$ can be perfectly estimated as
\begin{equation} 
  \bar{\mathbf{Q}}_{1} = c_{\theta}^{-1} \bar{\mathbf{Y}}_3  \mathbf{A}_3^H (\mathbf{A}_3 \mathbf{A}_3^H)^{-1} .  
\end{equation}

\subsection{Phase IV: Estimation of $\bar{\mathbf{B}}$} \label{Sec:IV-IV}

We can observe from \eqref{eq:yt_2} that the estimation of $\bar{\mathbf{B}}$ is challenging due to its coupling with $\bar{\mathbf{r}}_1$ and the interference term $\sum_{m'_1} \bar{r}_{1,k,m'_1} \bar{\mathbf{Q}}_{1} [\bm{\Phi}_{1,t}]_{:,m'_1}$, considering that the coefficients $\bar{r}_{1,k,m'_1}, \forall k,m'_1$ have not been estimated yet. 
To address these issues, we divide $\tau_4$ time instants into two parts with the same length $\tau_{4,1}$, i.e. $\tau_4 = 2 \tau_{4,1}$. 
At each time instant $t = \tau'_3 + \tau \in [\tau'_3 + 1, \tau'_3 +  \tau_{4,1}]$ of the first part, i.e., for each $\tau\in[\tau_{4,1}]$, all users are assigned a common, arbitrary scalar with unit power as their pilot signal, i.e.,
\begin{equation} \label{eq:x_B4}
  x_{k,\tau'_3 + \tau} = x_{\tau'_3 + \tau}, \;\; \forall k\in[K], \tau \in [\tau_{4,1}], 
\end{equation}
satisfying $|x_{\tau'_3 + \tau}|=1, \forall \tau \in [\tau_{4,1}]$. 
In this part, denote the scattering matrices of BD-RIS 1 and BD-RIS 2 as $\bm{\Phi}_{1,\tau'_3+\tau}$ and $\bm{\Phi}_{2,\tau'_3+\tau}, \forall \tau\in[\tau_{4,1}]$, respectively. 
In the second part, i.e., at each time instant $t = \tau'_3 + \tau_{4,1} + \tau \in [\tau'_3 + \tau_{4,1} + 1, \tau'_3 +  2\tau_{4,1}]$, we set the pilot signals and scattering matrices as follows
\begin{equation} \label{eq:IV2_x}
  x_{k,\tau'_3 + \tau_{4,1} + \tau} = x_{k,\tau'_3 + \tau} , \;\; \forall k\in[K], \tau\in[\tau_{4,1}] ,
\end{equation} 
\begin{align}  \label{eq:IV2_Psi}
  [\bm{\Phi}_{1,\tau'_3 + \tau_{4,1} + \tau}]_{:,m'_1}  = & \left\{ \!\!\!\begin{array}{ll} 
  e^{\jmath \theta} [\bm{\Phi}_{1,\tau'_3 + \tau}]_{:,1},\!\!\! & m'_1 = 1,  \\
  {[\bm{\Phi}_{1,\tau'_3 + \tau}]_{:,m'_1}} , & m'_1 \in [2, M_1], 
  \end{array} \right.   \notag\\
  &   \forall \tau\in[\tau_{4,1}] ,
\end{align}  
\begin{equation} \label{eq:IV2_Phi}
  \bm{\Phi}_{2,\tau'_3 + \tau_{4,1} + \tau} = \bm{\Phi}_{2,\tau'_3 + \tau}, \;\; \forall \tau\in[\tau_{4,1}] ,
\end{equation}    
where $\theta\in(0, 2\pi)$ is an arbitrary phase shift as aforementioned. 
Substituting the designed pilot signals and scattering matrices into \eqref{eq:yt_2} and removing the noise, we can obtain the received signals in the two parts, which are denoted as $\mathbf{y}_{\tau'_3 + \tau}$ and $\mathbf{y}_{\tau'_3 + \tau_{4,1} + \tau}, \forall \tau\in[\tau_{4,1}]$, respectively. Then, we can construct the effective received signal as follows 
\begin{align}  
  & \bar{\mathbf{y}}_{4,\tau} \!
  = \mathbf{y}_{\tau'_3 + \tau} - \mathbf{y}_{\tau'_3 + \tau_{4,1} + \tau} 
  - c_{\theta} x_{\tau'_3 + \tau} \bar{\mathbf{Q}}_{1} [\bm{\Phi}_{1, \tau'_3 + \tau}]_{:,1} \notag \\
  & = \!c_{\theta} x_{\tau'_3 + \tau}   \sum_{m_2=1}^{M_2} \sum_{m_1=1}^{M_1} \bar{b}_{m_2,m_1} \bar{\mathbf{Q}}_{2} [\bm{\Phi}_{1,\tau'_3 + \tau}]_{m_1,1} [\bm{\Phi}_{2,\tau'_3 + \tau}]_{:,m_2} \notag \\
  & =\! c_{\theta}   x_{\tau'_3 + \tau}  \!  \left(  [\bm{\Phi}_{1, \tau'_3 + \tau}]_{:,1}^T \!\otimes\! \left(  \bar{\mathbf{Q}}_{2} \bm{\Phi}_{2, \tau'_3 + \tau}  \right) \right) \!\bar{\mathbf{b}} ,   \;  \forall \tau \!\in\! [  \tau_{4,1}] , 
\end{align} 
where $\bar{\mathbf{b}} = \operatorname{vec}(\bar{\mathbf{B}})$ and $c_{\theta}$ is given in \eqref{eq:c_theta}. 
Then, the overall effective received signal is expressed as 
\begin{equation} 
  \bar{\mathbf{y}}_4 = [ \bar{\mathbf{y}}_{4, 1}^T , \ldots , \bar{\mathbf{y}}_{4, \tau_{4,1}}^T ]^T
  = c_{\theta}   \mathbf{A}_4  \bar{\mathbf{b}} , 
\end{equation}
where
\begin{equation}
  \mathbf{A}_4 = \begin{bmatrix}
   x_{\tau'_3 + 1} [\bm{\Phi}_{1,\tau'_3 + 1}]_{:,1}^T \otimes \left( \bar{\mathbf{Q}}_{2}  \bm{\Phi}_{2,\tau'_3 + 1} \right) \\
  \vdots \\
   x_{\tau'_3 + \tau_{4,1}} [\bm{\Phi}_{1,\tau'_3 + \tau_{4,1}}]_{:,1}^T \otimes \left( \bar{\mathbf{Q}}_{2} \bm{\Phi}_{2,\tau'_3 + \tau_{4,1}} \right)
  \end{bmatrix} 
  .  \label{eq:A4_2}
\end{equation}

In the following, we will introduce how to design $\bm{\Phi}_{1,\tau'_3 + \tau}$ and $\bm{\Phi}_{2,\tau'_3 + \tau}, \forall \tau\in[\tau_{4,1}]$, to make sure the matrix $\mathbf{A}_4$ has full column rank. 
Recalling that the SVD of $\bar{\mathbf{Q}}_{2}$ satisfies $\bar{\mathbf{Q}}_{2} = \mathbf{U}_{\bar{\mathbf{Q}}_2} \bm{\Sigma}_{\bar{\mathbf{Q}}_2} \mathbf{V}_{\bar{\mathbf{Q}}_2}^H$. Similar to \eqref{eq:phi_r2}, we design $\bm{\Phi}_{2,\tau'_3 + \tau}$~as 
\begin{equation} \label{eq:phi_B4}
  \bm{\Phi}_{2,\tau'_3 + \tau} = \mathbf{V}_{\bar{\mathbf{Q}}_2} \mathbf{P}_{\tau'_3 + \tau}, \;\; \forall \tau\in  [\tau_{4,1}], 
\end{equation}
where $\mathbf{P}_{\tau'_3 + \tau}$ is an $M_2\times M_2$ unitary matrix. 
The matrices $\mathbf{P}_{\tau'_3 + \tau}$ and $\bm{\Phi}_{1,\tau'_3 + \tau}$ are designed ad follows. 
Specifically, define $M_{1,0}$ as the smallest integer such that $\tau_0 = \frac{M_2M_{1,0}}{q_2}$ is an integer. Denote $\eta = \lfloor \frac{M_1}{M_{1,0}} \rfloor$. 
For each $\tau\in[\tau_{4,1}]$, we divide the $M_1$ elements in $[\bm{\Phi}_{1,\tau'_3 + \tau}]_{:,1}$ into $\eta+1$ groups, whose sizes are given by 
\begin{equation} 
  M_{1,n} = \left\{ \begin{array}{ll}
  M_{1,0}, & n \in [\eta] , \\
  M_1 - \eta M_{1,0} , & n = \eta + 1. 
  \end{array} \right.
\end{equation}
Denote the set of indexes in each of the $\eta+1$ groups as 
\begin{equation} 
  \mathcal{M}_{1,n} \!=\! \left\{ \!\!\! \begin{array}{ll} 
  \{ (n\!-\!1) M_{1,0}\! + \!1, \ldots, n M_{1,0} \}, &\!\! n \in [\eta] , \\
  \{ \eta M_{1,0} + 1, \ldots, M_1 \} , & \!\!n = \eta + 1.  
  \end{array} \right.
\end{equation}
Correspondingly, we divide the $\tau_{4,1}$ time instants into $\eta+1$ non-overlapping groups, with the size of group $n$ denoted as $T_n$, which is obtained by replacing $\tau_{2,1}$ in \eqref{eq:Ti_r2} with $\tau_{4,1}$. Denote the set of time instants in the $n$-th group as $\mathcal{T}_n$, which is obtained by changing $\tau_1$ and $\tau_{2,1}$ in \eqref{eq:Tiset_r2} to $\tau'_3$ and $\tau_{4,1}$, respectively.  
Then, $\forall n\in[\eta+1], r \in [T_n]$, the matrix $\mathbf{P}_{\mathcal{T}_n(r)}$ is designed as in \eqref{eq:P_r2}.  
For each $n \in [\eta+1]$ and $j \in [M_{1,n}]$, we define the set $\mathcal{R}_{n,j}$ as in \eqref{eq:Rij_r2}. 
Then, the first column of the scattering matrix of BD-RIS 1 is designed as
\begin{align} 
  & [\bm{\Phi}_{1,\mathcal{T}_{n'}(r)}]_{\mathcal{M}_{1,n}(j),1} \notag\\
  & = \left\{ \begin{array}{ll}
  \frac{1}{\!\sqrt{ \sum_{j'=1}^{M_{1,n}}1[ r \in \mathcal{R}_{n,j'} ] } } , & \text{if} \;\; n = n' \; {\rm{and}} \; r \in\mathcal{R}_{n,j}, \\
  0 , & \text{otherwise}, 
  \end{array} \right.  \notag\\
  &  \forall n\in [\eta+1], n' \in [\eta+1],  j\in[M_{1,n}], r\in [T_{n'}]. \label{eq:psi_B4} 
\end{align} 
For each $n' \in [\eta+1]$ and $r\in [T_{n'}]$, the remaining $M_1-1$ columns of $\bm{\Phi}_{1,\mathcal{T}_{n'}(r)}$ are constructed as an orthonormal basis for the orthogonal complement of its first column, thereby satisfying the constraint in \eqref{eq:Phi}.
Following the above criterion, in the case of $\tau_{4,1} = \lceil \frac{M_1M_2}{q_2} \rceil$ and $\tau_{4} = 2\lceil \frac{M_1M_2}{q_2} \rceil$, we can construct $\mathbf{A}_4$ with full column rank, i.e., $\operatorname{rank}(\mathbf{A}_4) = M_1M_2$. 
Then, $\bar{\mathbf{b}}$ can be perfectly estimated as
\begin{equation} 
  \bar{\mathbf{b}} =  c_{\theta}^{-1}  (\mathbf{A}_4^H \mathbf{A}_4)^{-1} \mathbf{A}_4^H \bar{\mathbf{y}}_4  .  
\end{equation}

\subsection{Phase V: Estimation of $\bar{\mathbf{r}}_{1}$} \label{Sec:IV-V}

At each time instant $t = \tau'_4 + \tau \in [\tau'_4 + 1, \tau'_4 +  \tau_5]$ of this phase, 
denote the pilot signal as $\mathbf{x}_{\tau'_4 + \tau}$, the scattering matrix of BD-RIS 1 as $\bm{\Phi}_{1,\tau'_4+\tau}$, the scattering matrix of BD-RIS 2 as $\bm{\Phi}_{2,\tau'_4+\tau}$, and the received signal as $\mathbf{y}_{\tau'_4 + \tau}$. 
Based on the estimated channels $\bar{\mathbf{Q}}_{2}$ and $\bar{\mathbf{r}}_2$ in the previous phases, we can obtain the effective received signal by removing the contributions made by the single-reflection link via BD-RIS 2 from $\mathbf{y}_{\tau'_4 + \tau}$, which is given by 
\begin{align}  
  \bar{\mathbf{y}}_{5,\tau}  \!
  & =\! \mathbf{y}_{\tau'_4+\tau} \! - \! \sum_{k=1}^{K} \! \sqrt{p} x_{k,\tau'_4+\tau} \!\sum_{m_2=1}^{M_2}  \! \bar{r}_{2,k,m_2} \bar{\mathbf{Q}}_{2} [\bm{\Phi}_{2,\tau'_4+\tau}]_{:,m_2}  \notag \\ 
  & =  \sqrt{p}   \left(  \mathbf{x}_{\tau'_4 + \tau}^T  \otimes \left(  \left( \bar{\mathbf{Q}}_{1}  
  + \bar{\mathbf{Q}}_{2}  \bm{\Phi}_{2, \tau'_4+\tau} \bar{\mathbf{B}} \right)   \bm{\Phi}_{1,\tau'_4+\tau}  \right) \right) 
   \notag\\
  & \;\;\; \times [ \bar{r}_{1,1,1} , \bar{\bm{r}}_{1}^T ]^T , \;\;  \forall \tau\in[\tau_5] .  \label{eq:bary5r1}  
\end{align} 
Then, by setting the scattering matrix of BD-RIS 2 to remain constant over $\tau_5$ time instants, i.e., $\bm{\Phi}_{2,\tau'_4+\tau} = \bm{\Phi}_2, \forall \tau\in[\tau_5]$, we can express the overall effective received signal as
\begin{equation} \label{eq:bary5r1_all} 
  \bar{\mathbf{y}}_5 = [ \bar{\mathbf{y}}_{5, 1}^T , \ldots , \bar{\mathbf{y}}_{5, \tau_{5}}^T ]^T
  = \sqrt{p}  \mathbf{A}_5  [ \bar{r}_{1,1,1} , \bar{\bm{r}}_{1}^T ]^T  , 
\end{equation}
where 
\begin{equation}
  \mathbf{A}_5 = \begin{bmatrix}
   \mathbf{x}_{\tau'_4 + 1}^T  \otimes \mathbf{F} \bm{\Phi}_{1,\tau'_4+1}  \\
  \vdots \\
   \mathbf{x}_{\tau'_4 + \tau_5}^T  \otimes \mathbf{F}  \bm{\Phi}_{1,\tau'_4+\tau_5} 
  \end{bmatrix} , \label{eq:A5_2}
\end{equation} 
\begin{equation}\label{eq:V_Ft}
  \mathbf{F} = \bar{\mathbf{Q}}_{1} + \bar{\mathbf{Q}}_{2}  \bm{\Phi}_2 \bar{\mathbf{B}}  . 
\end{equation}

To characterize the number of time instants required for perfect estimation of $\bar{\mathbf{r}}_1$, we first analyze the rank of the matrix $\mathbf{F}$. 
Specifically, let us express the SVD of the matrix $\bar{\mathbf{Q}}_{1}$ as $\bar{\mathbf{Q}}_{1 } = \mathbf{U}_{\bar{\mathbf{Q}}_{1 }} \bm{\Sigma}_{\bar{\mathbf{Q}}_{1 }} \mathbf{V}_{\bar{\mathbf{Q}}_{1 }}^H$, where $\mathbf{U}_{\bar{\mathbf{Q}}_{1}}  \in \mathbb{C}^{L \times L} $ and $\mathbf{V}_{\bar{\mathbf{Q}}_{1 }}  \in \mathbb{C}^{M_1\times M_1}$ are unitary matrices, and $ \bm{\Sigma}_{\bar{\mathbf{Q}}_{1 }} =  \operatorname{blkdiag} \{ \bm{\Lambda}_{\bar{\mathbf{Q}}_{1 }} , \bm{0}_{L-q_1,M_1-q_1}\}$ is an $L\times M_1$ rectangular diagonal matrix with $\bm{\Lambda}_{\bar{\mathbf{Q}}_{1 }} = \operatorname{diag}\{\sigma_{\bar{\mathbf{Q}}_{1 },1} , \ldots, \sigma_{\bar{\mathbf{Q}}_{1 },q_1} \}$ containing $q_1$ nonzero singular values. 
Let us partition the unitary matrices as $\mathbf{U}_{\bar{\mathbf{Q}}_1} = [\mathbf{U}_{\bar{\mathbf{Q}}_1, 1}, \mathbf{U}_{\bar{\mathbf{Q}}_1, 2}]$ and $\mathbf{V}_{\bar{\mathbf{Q}}_1} = [\mathbf{V}_{\bar{\mathbf{Q}}_1, 1}, \mathbf{V}_{\bar{\mathbf{Q}}_1, 2}]$. The submatrices $\mathbf{U}_{\bar{\mathbf{Q}}_1, 1} \in \mathbb{C}^{L \times q_1}$ and $\mathbf{V}_{\bar{\mathbf{Q}}_1, 1} \in \mathbb{C}^{M_1 \times q_1}$ consist of the singular vectors corresponding to the non-zero singular values, while $\mathbf{U}_{\bar{\mathbf{Q}}_1, 2} \in \mathbb{C}^{L \times (L-q_1)}$ and $\mathbf{V}_{\bar{\mathbf{Q}}_1, 2} \in \mathbb{C}^{M_1 \times (M_1-q_1)}$ correspond to the zero singular values. 
Then, in Theorem~\ref{Theorem:rank}, we characterize the maximum rank of $\mathbf{F}$ via the optimal design of the scattering matrix $\bm{\Phi}_2$. 
\begin{Theorem} \label{Theorem:rank} 
Given the channel matrices $\bar{\mathbf{Q}}_{1} , \bar{\mathbf{Q}}_{2},$ and $\bar{\mathbf{B}}$, the maximum rank of $\mathbf{F}$ achievable via the configuration of the scattering matrix $\bm{\Phi}_2$ is given by
\begin{subequations}
\begin{align}
  f & = \max_{\bm{\Phi}_2} \;\operatorname{rank}(\mathbf{F})   \\
  & = \min \!\left\{  \operatorname{rank}\!\left( [ \bar{\mathbf{Q}}_{1} , \bar{\mathbf{Q}}_{2} ]  \right) ,  
  \operatorname{rank}\! \left( \begin{bmatrix}
                 \bar{\mathbf{Q}}_{1} \\
                 \bar{\mathbf{B}} 
               \end{bmatrix} \right) \right\}\!.  \label{eq:V_B_rankF}
\end{align}
\end{subequations}
This maximum rank is achieved by designing $\bm{\Phi}_2$ as 
\begin{equation}\label{eq:V_B_phi}
  \bm{\Phi}_2 = e^{\jmath \phi} \mathbf{V}_{\tilde{\mathbf{Q}}_2} \mathbf{U}_{\tilde{\mathbf{B}}}^H, 
\end{equation} 
where $\mathbf{V}_{\tilde{\mathbf{Q}}_2} \in \mathbb{C}^{M_2\times M_2}$ is the right singular matrix of the matrix $\tilde{\mathbf{Q}}_{2}  = \mathbf{U}_{\bar{\mathbf{Q}}_{1},2}^H \bar{\mathbf{Q}}_{2}$, 
$\mathbf{U}_{\tilde{\mathbf{B}}} \in \mathbb{C}^{M_2\times M_2}$ is the left singular matrix of $\tilde{\mathbf{B}} = \bar{\mathbf{B}} \mathbf{V}_{\bar{\mathbf{Q}}_{1},2}$, 
and the phase $\phi$ satisfies
\begin{equation}\label{eq:V_B_phi_phi}
  \phi \in [0, 2\pi) \setminus \tilde{\Theta} .  
\end{equation}
The set $\tilde{\Theta}$ is given by
\begin{equation}\label{eq:V_B_phi_Thetatilde}
  \tilde{\Theta} \!=\! \left\{ \left. \arg\!\left(- \tfrac{1}{\lambda_{\tilde{\mathbf{F}}, i }} \right) \; \operatorname{mod} \; 2\pi \;\! \right| \;\! i \in [q_1] ,  |\lambda_{\tilde{\mathbf{F}}, i }| = 1 \right\}, 
\end{equation}
where $\lambda_{\tilde{\mathbf{F}}, 1}, \ldots, \lambda_{\tilde{\mathbf{F}}, q_1}$ denote the eigenvalues of $\tilde{\mathbf{F}}$ given by 
\begin{equation} \label{eq:V_Ftilde}
  \tilde{\mathbf{F}} \!\!=\! \bm{\Lambda}_{ \bar{\mathbf{Q}}_{1 }}^{\! -1}  \! \mathbf{U}_{\! \bar{\mathbf{Q}}_1 ,1}^H   \bar{\mathbf{Q}}_{2} \!  \mathbf{V}_{ \!\tilde{\mathbf{Q}}_2} 
  \!\!\operatorname{blkdiag}\{ \bm{0}_{c,c} , \!\mathbf{I}_{M_2\!-c} \}  \mathbf{U}_{\tilde{\mathbf{B}} }^H   \bar{\mathbf{B}}  \mathbf{V}_{ \!\bar{\mathbf{Q}}_1 ,1}, 
\end{equation}
with $c = \min\{ \operatorname{rank}(\tilde{\mathbf{Q}}_{2}) , \operatorname{rank}(\tilde{\mathbf{B}}) \} $. 
\begin{IEEEproof}
See Appendix A. 
\end{IEEEproof}
\end{Theorem}

Theorem~\ref{Theorem:rank} characterizes the maximum rank $f$ of $\mathbf{F}$ for given channel realizations $\bar{\mathbf{Q}}_{1}, \bar{\mathbf{Q}}_{2},$ and $\bar{\mathbf{B}}$ via the design of $\bm{\Phi}_2$. 
In the Corollary \ref{Corollary:rank}, we establish the upper and lower bounds of $f$ across all possible channel realizations $\bar{\mathbf{Q}}_{1}, \bar{\mathbf{Q}}_{2},$ and $\bar{\mathbf{B}}$.  
\begin{Corollary} \label{Corollary:rank} 
  The upper bound of $f$ is given by
  \begin{subequations}
  \begin{align} 
    f &\leq \max_{\bar{\mathbf{Q}}_{1}, \bar{\mathbf{Q}}_{2}, \bar{\mathbf{B}} } \max_{\bm{\Phi}_2} \;\operatorname{rank}(\mathbf{F})  \\
      & = \min \{  q_1 + \min\{ q_2 , b \} , L , M_1 \}.  \label{eq:V_B_rankF_upper}
  \end{align}
  \end{subequations}
  The lower bound of $f$ is given by
  \begin{subequations}
  \begin{align}
    f &\geq \min_{\bar{\mathbf{Q}}_{1}, \bar{\mathbf{Q}}_{2}, \bar{\mathbf{B}} } \max_{\bm{\Phi}_2} \;\operatorname{rank}(\mathbf{F}) \\
      & = \max\{ q_1 , \min\{q_2, b\} \}.  \label{eq:V_B_rankF_lower}
  \end{align}
  \end{subequations}
\begin{IEEEproof}
See Appendix B. 
\end{IEEEproof}
\end{Corollary}

We can observe from \eqref{eq:V_B_rankF} that the rank $f$ is fundamentally determined by the geometric relationship among the subspaces spanned by $\bar{\mathbf{Q}}_{1}$, $\bar{\mathbf{Q}}_{2}$, and $\bar{\mathbf{B}}$. 
The upper bound in \eqref{eq:V_B_rankF_upper} corresponds to the best-case scenario where the links BD-RIS 1 - BS and BD-RIS 1 - BD-RIS 2 - BS  provide distinct spatial directions. This maximum rank is achieved when both the intersection between the column spaces of $\bar{\mathbf{Q}}_{1}$ and $\bar{\mathbf{Q}}_{2}$ and that between the row spaces of $\bar{\mathbf{Q}}_{1}$ and $\bar{\mathbf{B}}$ are trivial , i.e., $\dim (\operatorname{Col}(\bar{\mathbf{Q}}_{1}) \cap \operatorname{Col}(\bar{\mathbf{Q}}_{2}) )= 0$ and $\dim (\operatorname{Row}(\bar{\mathbf{Q}}_{1}) \cap \operatorname{Row} (\bar{\mathbf{B}}) )= 0$ \cite{ref:Matrix}. 
In contrast, the lower bound in \eqref{eq:V_B_rankF_lower} corresponds to the worst-case scenario in which the channel subspaces are maximally aligned. Specifically, when $\operatorname{Col}(\bar{\mathbf{Q}}_{2}) \subseteq \operatorname{Col}(\bar{\mathbf{Q}}_{1})$ or $\operatorname{Row}(\bar{\mathbf{B}}) \subseteq \operatorname{Row}(\bar{\mathbf{Q}}_{1})$, the rank is limited by the BD-RIS 1 - BS channel $\bar{\mathbf{Q}}_{1}$ and thus reduces to $q_1$. 
On the other hand, the cascaded link BD-RIS 1 - BD-RIS 2 - BS becomes the bottleneck when $\bar{\mathbf{Q}}_{1}$ spans a smaller subspace. Specifically, the rank reduces to $q_2$ in the case of $\operatorname{Col}(\bar{\mathbf{Q}}_{1}) \subseteq \operatorname{Col}(\bar{\mathbf{Q}}_{2})$ and $q_2 < b$, whereas it reduces to $b$ in the case of $\operatorname{Row}(\bar{\mathbf{Q}}_{1}) \subseteq \operatorname{Row}(\bar{\mathbf{B}})$ and $b \leq q_2$.

After designing the scattering matrix $\bm{\Phi}_2$ of BD-RIS 2 and characterizing the rank of $\mathbf{F}$, we express its SVD as $\mathbf{F} = \mathbf{U}_{\mathbf{F}} \bm{\Sigma}_{\mathbf{F}} \mathbf{V}_{\mathbf{F}}^H$, where $\mathbf{U}_{\mathbf{F}} \in\mathbb{C}^{L\times L}$ and $\mathbf{V}_{\mathbf{F}} \in\mathbb{C}^{M_1\times M_1}$ are unitary matrices, and $\bm{\Sigma}_{\mathbf{F}} =  \operatorname{blkdiag} \{ \bm{\Lambda}_{\mathbf{F} } , \bm{0}_{L-f,M_1-f}\}$ is an $L\times M_1$ rectangular diagonal matrix with $\bm{\Lambda}_{\mathbf{F} } = \operatorname{diag}\{\sigma_{\mathbf{F},1} , \ldots, \sigma_{\mathbf{F},f} \}$ containing $f$ nonzero singular values. 
Then, we design the scattering matrix of BD-RIS 1 as 
\begin{equation} \label{eq:V_r1_psi}
  \bm{\Phi}_{1,\tau'_4 + \tau} = \mathbf{V}_{\mathbf{F}} \mathbf{D}_{\tau'_4 + \tau}, \;\; \forall \tau\in[\tau_5], 
\end{equation}
where $\mathbf{D}_{\tau'_4 + \tau}$ is an $M_1\times M_1$ unitary matrix. Considering that the matrix $\mathbf{A}_5$ in \eqref{eq:A5_2} has a similar structure as $\mathbf{A}_2$ in \eqref{eq:A2}, we can follow similar lines as in Section \ref{Sec:IV-II} to design $\mathbf{x}_{\tau'_4 + \tau}$ and $\mathbf{D}_{\tau'_4 + \tau}, \forall \tau\in[\tau_5]$. In this way, for given channel matrices $\bar{\mathbf{Q}}_{1} , \bar{\mathbf{Q}}_{2},$ and $\bar{\mathbf{B}}$, we can construct $\mathbf{A}_5$ with full column rank, i.e., $\operatorname{rank}(\mathbf{A}_5) = KM_1$, in the case of $\tau_{5} = \lceil \frac{KM_1}{ f } \rceil$. 
Then, 
$\bar{\bm{r}}_{1}$ can be perfectly estimated as
\begin{equation} 
  \bar{\bm{r}}_{1} = ( \sqrt{p} )^{-1} \left[ (\mathbf{A}_5^H \mathbf{A}_5)^{-1} \right]_{2:KM_1,:} \mathbf{A}_5 \bar{\mathbf{y}}_5   .  
\end{equation}
According to Corollary~\ref{Corollary:rank}, for arbitrary channel realizations $\bar{\mathbf{Q}}_{1}, \bar{\mathbf{Q}}_{2}$, and $\bar{\mathbf{B}}$, including the worst ones, with fixed ranks $q_1, q_2$, and $b$, respectively, it is sufficient to use $\tau_{5} = \lceil \frac{KM_1}{ \max\{ q_1 , \min\{q_2, b\} \} } \rceil$ time instants to guarantee the perfect estimation of $\bar{\bm{r}}_{1}$ based on the proposed scheme. In contrast, under the most favorable channel conditions, the number of times instants can be reduced to $\tau_{5} = \lceil \frac{KM_1}{ \min\{ q_1 + \min\{ q_2 , b \} , L , M_1 \} } \rceil$.

\subsection{Overall Overhead}


In the double-BD-RIS cooperatively assisted multi-user MIMO communication system without noise, the proposed framework achieves perfect estimation of the channel matrices $\bar{\mathbf{Q}}_{1}, \bar{\mathbf{Q}}_{2}, \bar{\bm{r}}_{1}, \bar{\bm{r}}_{2}$, and $\bar{\mathbf{B}}$ with a pilot overhead of
\begin{equation} \label{eq:overhead}
  T_{\text {double-BD-RIS}} =  4M_2 + 2 \left\lceil \tfrac{KM_2}{q_2} \right\rceil + 4M_1 +  2\left\lceil \tfrac{M_1M_2}{q_2} \right\rceil + \left\lceil \tfrac{KM_1}{ f } \right\rceil  , 
\end{equation} 
where $q_2$ denotes the rank of $\bar{\mathbf{Q}}_{2}$, and $f$ corresponds to the rank of the channel matrix aggregated by the BD-RIS 1 - BS and BD-RIS 1 - BD-RIS 2 - BS links as shown in \eqref{eq:V_B_rankF}. 
Based on the above estimates, the cascaded channels $\mathbf{J}_{1,k}, \mathbf{J}_{2,k}$, and $\mathbf{J}_{1,2,k}, \forall k\in[K]$, which consist of $KL(M_1^2 + M_2^2 + M_1^2M_2^2)$ elements, can be perfectly recovered applying \eqref{eq:J1k}, \eqref{eq:J12k}, \eqref{eq:Q1kn_2}, \eqref{eq:barr1kn11}, and \eqref{eq:Q12kmnn_2}.

It is evident that by exploiting the intrinsic channel correlation properties in double-BD-RIS-aided multi-user MIMO networks, the overhead of our proposed scheme is of a much smaller order than that required by treating all channel entries as independent elements and directly estimating the cascaded channels, where an overhead of $K(M_1^2 + M_2^2 + M_1^2M_2^2)$ time instants is required. 
Particularly, as shown in \eqref{eq:overhead}, our approach eliminates the prohibitive quadratic dependence on the BD-RIS dimensions (i.e., $M_1^2$ and $M_2^2$) and avoids the product of $K, M_1^2$, and $M_2^2$. 
Instead, the overhead of our approach is dominated by pairwise products of three parameters, i.e., $KM_2, M_1M_2$, and $KM_1$, normalized by corresponding ranks, thereby drastically reducing the required training overhead.

In the following, we compare the training overhead of the considered double-BD-RIS-aided system with that of the double-diagonal-RIS, single-BD-RIS, and single-diagonal-RIS counterparts. 
\begin{Remark}
In the double-diagonal-RIS scenario, a time duration consisting of 
\begin{equation} \label{eq:overhead_double_D_RIS}
  T_{\text {double-D-RIS}} \!=\! M_1 + M_2 + \left\lceil \tfrac{(K-1)M_1}{ q_1 } \right\rceil + \left\lceil \tfrac{(K-1)M_2}{q_2} \right\rceil + \left\lceil \tfrac{M_1M_2}{q_2} \right\rceil \! , 
\end{equation}                
pilot symbols is sufficient for perfect channel estimation~\cite{ref:doubleRIS_ZhengBY_conf}.  
The comparison between \eqref{eq:overhead} and \eqref{eq:overhead_double_D_RIS} shows that by utilizing the channel properties to reduce the number of independent variables to estimate, the training overhead of the double-BD-RIS network can be on the same order as that of the double-diagonal-RIS network, although the number $KL(M_1^2 + M_2^2 + M_1^2M_2^2)$ of channel coefficients in the double-BD-RIS network is significantly larger than that of the double-diagonal-RIS network, given by $KL(M_1 + M_2 + M_1M_2)$. 
Thus, the rate gain arising from designing two non-diagonal scattering matrices can be achieved without incurring a substantially higher estimation cost than that of double-diagonal-RIS systems. 
\end{Remark}

\begin{Remark}
Consider the scenario where the numbers of elements of two BD-RISs are on the same order, i.e., $M_1 = \Theta(M)$ and $M_2 = \Theta(M)$, and the channels exhibit rich scattering such that the ranks of both $\bar{\mathbf{Q}}_{1}$ and $\bar{\mathbf{Q}}_{2}$ scale with the BD-RIS dimensions, yielding $f = \Theta(M)$ according to \eqref{eq:V_B_rankF}. Then, the training overhead given in \eqref{eq:overhead} for perfect channel recovery in double-BD-RIS systems is in the order of 
\begin{equation} \label{eq:overhead2}
  T_{\text {double-BD-RIS}} = \Theta(M+K).
\end{equation} 
In the scenario with a single $M_1$-element BD-RIS under rich scattering conditions, the training overhead is given by \cite{ref:wangrui}
\begin{equation}  \label{eq:overhead_single_BD_RIS}
  T_{\text {single-BD-RIS}} = 2M_1 + \left\lceil \tfrac{M_1(K-1)}{q_1} \right\rceil = \Theta(M+K). 
\end{equation}
Thus, although the number of channel coefficients of the double-BD-RIS network is significantly larger than that of the single-BD-RIS network, i.e., $KL(M_1^2 + M_2^2 + M_1^2M_2^2)$ versus $KLM_1^2$, the overhead of the double-BD-RIS system can be reduced to the same order as that of the single-BD-RIS system, both scaling linearly with the number of RIS elements and users. This indicates that the substantial performance gain enabled by cooperative beamforming of two BD-RISs can be realized at a channel estimation cost comparable to that of the single-BD-RIS counterpart. 
\end{Remark}

\begin{Remark} 
Consider the scenario with a single $M_1$-element diagonal RIS satisfying $M_1 = \Theta(M)$, where the number of channel coefficients is $KLM_1$. Under rich scattering conditions, the training overhead is given by \cite{ref:WangZR}
\begin{equation}  \label{eq:overhead_single_D_RIS}
  T_{\text {single-D-RIS}} = M_1 + \left\lceil \tfrac{M_1(K-1)}{q_1} \right\rceil = \Theta(M+K). 
\end{equation}
The comparison between \eqref{eq:overhead_single_BD_RIS} and \eqref{eq:overhead_single_D_RIS} verifies the theoretical performance gain of double-BD-RIS-assisted communication over single-diagonal-RIS counterpart even with channel estimation overhead considered.
\end{Remark}


\section{Channel Estimation in the Case with Noise} \label{Sec:CE_AWGN}

Building on the low-overhead channel estimation framework for the ideal noiseless case introduced before, this section focuses on how to estimate the cascaded channels in the practical scenario with noise at the BS.

\subsection{Phase I: Estimation of $\bar{\mathbf{Q}}_{2}$} \label{Sec:V-I}

To estimate $\bar{\mathbf{Q}}_{2}$ in the presence of noise at the BS, we divide the total duration $\tau_1$ into four parts of equal length $\tau_{1,1}$, where $\tau_{1}$ is larger than the minimum required overhead $4M_2$ in the ideal noiseless case to enhance estimation performance. 
At each time instant $t = \tau \in [\tau_{1,1}]$ of the first part, we design the pilot signal $x_{k,\tau}$ of user $k$ and the scattering matrix $\bm{\Phi}_{1,\tau}$ of BD-RIS 1 according to \eqref{eq:I1_x} and \eqref{eq:I1_Psi}, respectively. For BD-RIS 2, the scattering matrix $\bm{\Phi}_{2,\tau}$ is designed following \eqref{eq:I1_Phi} for $\tau \in [M_2]$, while being a randomly generated unitary matrix for each $\tau \in [M_2+1, \tau_{1,1}]$.  
In the subsequent three parts, the pilot signals and scattering matrices for both BD-RISs are designed as in \eqref{eq:I2_x}-\eqref{eq:I4_Phi}. 
Then, similar to \eqref{eq:bary1}, the overall effective received signal in the noisy case is given by
\begin{equation} \label{eq:Y1_noise}
  \bar{\mathbf{Y}}_1 = c_{\theta} \bar{\mathbf{Q}}_{2} \mathbf{A}_1 +  \mathbf{W}_{1} , 
\end{equation}
where 
$\mathbf{A}_1 \!=\! [ x_{1} [\bm{\Phi}_{2,1}]_{:,1} , \ldots , x_{\tau_{1,1}} [\bm{\Phi}_{2,\tau_{1,1}}]_{:,1} ]$ 
and $\mathbf{W}_{1} = [\mathbf{w}_{1,1}, \ldots, \mathbf{w}_{1,\tau_{1,1}}]$ 
with $\mathbf{w}_{1,\tau} = \frac{ \mathbf{z}_{\tau} + \mathbf{z}_{\tau_{1,1}+\tau} }{2} - \frac{\mathbf{z}_{2\tau_{1,1}\!+\tau} + \mathbf{z}_{3\tau_{1,1}+\tau}}{2} \!\!\sim\! \mathcal{CN}\!\left( \bm{0}, \sigma^2 \mathbf{I}_L \right)$. 
Then, we apply the LS estimator to estimate $ \bar{\mathbf{Q}}_{2}$ as follows 
\begin{equation} 
  \hat{\bar{\mathbf{Q}}}_{2}  = c_{\theta}^{-1} \bar{\mathbf{Y}}_1  \mathbf{A}_1^H (\mathbf{A}_1 \mathbf{A}_1^H)^{-1} .  
\end{equation}

In this work, we treat the linear combination of the channels from all users to an element of BD-RIS $i$ and then to the BS as the reference channel of the single-reflection link via BD-RIS $i$, as shown in \eqref{eq:Q1kn_2_c1} and \eqref{eq:barQ1}. Under this formulation, all users are allowed to transmit pilot sequences simultaneously, as shown in \eqref{eq:I1_x}. 
In contrast, the work \cite{ref:wangrui} adopted another approach, where the scalar $c_i$ in \eqref{eq:Q1kn_2_c1} is set to be $r_{i,1,1}, \forall i$. In this way, the cascaded channels are recovered by estimating one typical user's channel and a set of coefficients associated with the remaining users. During the estimation of the typical user's channel, other users are required to remain silent. Based on this approach, the overall effective received signal in \eqref{eq:Y1_noise} becomes
\begin{equation} \label{eq:Y1_noise2}
  \bar{\mathbf{Y}}_1 = c_{\theta} \tfrac{r_{1,1,1}}{\sum_{k=1}^{K} r_{i,k,1}} \bar{\mathbf{Q}}_{2} \mathbf{A}_1 +  \mathbf{W}_{1}. 
\end{equation}
A comparison between \eqref{eq:Y1_noise2} and \eqref{eq:Y1_noise} reveals that by allowing all users to transmit pilots simultaneously, our formulation contributes to higher received SNR in practical noisy environments than the approach in \cite{ref:wangrui}.

\subsection{Phase II: Estimation of $\bar{\bm{r}}_{2}$} \label{Sec:V-II}

We divide the total duration $\tau_2$ into two parts with the same length $\tau_{2,1}$. 
In the noiseless case, it is sufficient to perfectly recover $\bar{\mathbf{r}}_2$ with $\tau_{2,1} = \lceil \frac{KM_2}{q_2} \rceil$. 
In the noisy case, $\tau_{2,1}$ should be larger than $\lceil \frac{KM_2}{q_2} \rceil$ to improve estimation performance. 
During $\tau_{2,1}$ time instants of the first part, the scattering matrix of BD-RIS 1 is generated based on \eqref{eq:psi_r2}. 
For the initial $\lceil \frac{KM_2}{q_2} \rceil$ time instants of the first part, i.e., for each $\tau \in [\lceil \frac{KM_2}{q_2} \rceil]$, we set the pilot signal $\mathbf{x}_{\tau_1 + \tau}$ as in \eqref{eq:x_r2} and scattering matrix $\bm{\Phi}_{2,\tau_1 + \tau}$ of BD-RIS 2 as in \eqref{eq:phi_r2} and \eqref{eq:P_r2} with the difference of replacing $\bar{\mathbf{Q}}_{2}$ with its estimate $\hat{\bar{\mathbf{Q}}}_{2}$. 
For the remaining time instants of the first part, we generate $x_{k,\tau_1 + \tau}$'s as arbitrary unit-power scalars and $\bm{\Phi}_{2,\tau_1 + \tau}$ as a random unitary matrix, $\forall \tau \in [\lceil \frac{KM_2}{q_2} \rceil+1, \tau_{2,1}]$. 
In the second part, the pilot signals of $K$ users and both scattering matrices are designed as in \eqref{eq:II2_x} - \eqref{eq:II2_Phi}. 
Then, the effective received signal in the noisy case is expressed as 
\begin{equation} 
  \bar{\mathbf{y}}_2 = \!\sqrt{p} \hat{\mathbf{A}}_2  [ \bar{r}_{2,1,1},  \bar{\bm{r}}_{2}^T]^T  + \!\sqrt{p} ( \mathbf{A}_2 \!-\! \hat{\mathbf{A}}_2 ) [ \bar{r}_{2,1,1},  \bar{\bm{r}}_{2}^T]^T  + \mathbf{w}_2  , \label{eq:bary2_noise}
\end{equation}
where $\mathbf{w}_2 \!=\! \left[\! \frac{( \mathbf{z}_{\tau_1 + 1}\! + \mathbf{z}_{\tau_1 + \tau_{2,1} + 1} )^T}{2}\!, \ldots, \!\frac{ (\mathbf{z}_{\tau_1 + \tau_{2,1}} \!+ \mathbf{z}_{\tau_1 + 2\tau_{2,1}} )^T}{2} \right]^{\!T} \!\!\sim \mathcal{CN}\left(0,\frac{\sigma^2}{2}\mathbf{I}_{L\tau_{2,1}}\right)$, 
$\mathbf{A}_2$ is given in \eqref{eq:A2}, and $\hat{\mathbf{A}}_2$ is obtained by replacing $\bar{\mathbf{Q}}_{2}$ in \eqref{eq:A2} with $\hat{\bar{\mathbf{Q}}}_{2}$. 
We can increase the pilot length $\tau_{1}$ in Phase I such that the error propagated from Phase I becomes negligible, i.e., $\mathbf{A}_2 - \hat{\mathbf{A}}_2 \to \bm{0}$. Then, we have
\begin{equation} 
  \bar{\mathbf{y}}_2 
  \approx \sqrt{p} \hat{\mathbf{A}}_2  [ \bar{r}_{2,1,1},  \bar{\bm{r}}_{2}^T]^T  + \mathbf{w}_2  . \label{eq:bary2_noise_approx}
\end{equation}
Applying the LS criterion, the vector $\bar{\bm{r}}_{2}$ can be estimated as 
\begin{equation} 
  \hat{\bar{\bm{r}}}_{2} = ( \sqrt{p} )^{-1} [ (  \hat{\mathbf{A}}_2^H \hat{\mathbf{A}}_2 )^{-1} ]_{2:KM_2,:} \hat{\mathbf{A}}_2^H \bar{\mathbf{y}}_2  .  
\end{equation}

\subsection{Phase III: Estimation of $\bar{\mathbf{Q}}_{1}$} \label{Sec:V-III}

To estimate $\bar{\mathbf{Q}}_{1}$, 
we divide the total duration $\tau_3$ into four parts with the same length $\tau_{3,1}$, where $\tau_3$ is larger than the minimum required overhead $4M_1$ in the noiseless case. 
At each time instant $t = \tau'_2 + \tau \in [\tau'_2 + 1, \tau'_2 + \tau_{3,1}]$ of the first part, we design the pilot signal $x_{k,\tau'_2+\tau}$ of user $k$ 
and the scattering matrix $\bm{\Phi}_{2,\tau'_2+\tau}$ of BD-RIS 2 as in Section~\ref{Sec:IV-III}. 
For BD-RIS 1, the scattering matrix $\bm{\Phi}_{1,\tau'_2+\tau}$ is designed as in Section~\ref{Sec:IV-III} for $\tau \in [M_1]$, while randomly generated for $\tau \in [M_1+1, \tau_{3,1}]$.  
In the subsequent three parts, the pilot signals and scattering matrices for both BD-RISs are designed following the rule in Section \ref{Sec:IV-III}. 
Then, in the noisy scenario, the overall effective received signal \eqref{eq:yt_noAWGN_singleUE_Q111_all} becomes
\begin{equation} \label{eq:bary3_noise} 
  \bar{\mathbf{Y}}_3 = c_{\theta} \bar{\mathbf{Q}}_{1} \mathbf{A}_3 + \mathbf{W}_3, 
\end{equation}
where $\mathbf{A}_3 = [ x_{\tau'_2 +1} [\bm{\Phi}_{1,\tau'_2 + 1}]_{:,1} , \ldots , x_{\tau'_2 + \tau_{3,1}} [\bm{\Phi}_{1,\tau'_2 + \tau_{3,1}}]_{:,1} ]$ 
and $\mathbf{W}_3 = [\mathbf{w}_{3,1}, \ldots, \mathbf{w}_{3,\tau_{3,1}}]$  
with $\mathbf{w}_{3,t} \sim \mathcal{CN}\left( \bm{0}, \sigma^2 \mathbf{I}_L \right)$. 
Similar to Phase I, since all users are allowed to transmit pilots for estimating the reference channel of the single refection link via BD-RIS 1, our approach yields higher received SNR than the method used in \cite{ref:wangrui}, where $K-1$ users are required to remain silent.  
Then, we apply the LS estimator to estimate $ \bar{\mathbf{Q}}_{1}$ as follows 
\begin{equation} 
  \hat{\bar{\mathbf{Q}}}_{1}  = c_{\theta}^{-1} \bar{\mathbf{Y}}_3  \mathbf{A}_3^H (\mathbf{A}_3 \mathbf{A}_3^H)^{-1} .  
\end{equation}

\subsection{Phase IV: Estimation of $\bar{\mathbf{B}}$} \label{Sec:V-IV}

We divide $\tau_4$ time instants into two parts with the same length $\tau_{4,1}> \lceil \tfrac{M_1M_2}{q_2} \rceil$. 
During the first part, we generate the pilot signal $x_{k,\tau'_3 + \tau}, \forall k\in[K], \tau \in [\tau_{4,1}]$ based on \eqref{eq:x_B4}. 
For the initial $\lceil \frac{M_1M_2}{q_2} \rceil$ time instants of the first part, i.e., for each $\tau \in [\lceil \frac{M_1M_2}{q_2} \rceil]$, we set the scattering matrices $\bm{\Phi}_{2,\tau'_3 + \tau}$ and $\bm{\Phi}_{1,\tau'_3 + \tau}$ as in Section~\ref{Sec:IV-IV} with the difference of replacing $\bar{\mathbf{Q}}_{2}$ with $\hat{\bar{\mathbf{Q}}}_{2}$. 
In the second part, the pilots and scattering matrices of both BD-RISs are designed as in \eqref{eq:IV2_x} - \eqref{eq:IV2_Phi}. 
Then, the overall effective received signal in the noisy case is expressed as 
\begin{equation} 
  \bar{\mathbf{y}}_4 
  =  c_{\theta} \hat{\mathbf{A}}_4  \bar{\mathbf{b}} + c_{\theta} ( \mathbf{A}_4 - \hat{\mathbf{A}}_4 )  \bar{\mathbf{b}}
  + \mathbf{n}_{4} + \mathbf{w}_{4}, 
\end{equation}
where $\bar{\mathbf{b}} = \operatorname{vec}(\bar{\mathbf{B}})$, $\mathbf{A}_4$ is given in \eqref{eq:A4_2}, $\hat{\mathbf{A}}_4$ is obtained by replacing the matrix $\bar{\mathbf{Q}}_{2}$ in \eqref{eq:A4_2} with its estimate $\hat{\bar{\mathbf{Q}}}_{2}$, 
$\mathbf{n}_{4} = [ {\mathbf{n}}_{4, 1}^T , \ldots , {\mathbf{n}}_{4, \tau_{4,1}}^T ]^T$ with $\mathbf{n}_{4,\tau} = c_{\theta} x_{\tau'_3 + \tau}  ( \bar{\mathbf{Q}}_{1} - \hat{\bar{\mathbf{Q}}}_{1} ) [\bm{\Phi}_{1,\tau'_3 + \tau}]_{:,1}$, 
and $\mathbf{w}_{4} = [ (\mathbf{z}_{\tau'_3 + 1} \!-\! \mathbf{z}_{\tau'_3 + \tau_{4,1} + 1})^{\!T} \!, \ldots ,\! (\mathbf{z}_{\tau'_3 + \tau_{4,1}} \!\! -  \mathbf{z}_{\tau'_3 + 2\tau_{4,1} })^T ]^T \!\sim\! \mathcal{CN}\!\left( \bm{0}, 2\sigma^2\mathbf{I}_{L\tau_{4,1}}\right)$. 
By increasing $\tau_{1}$ and $\tau_{3}$, the errors propagated from Phase I and Phase III become negligible, i.e., $\mathbf{A}_4 - \hat{\mathbf{A}}_4 \to \bm{0}$ and $\bar{\mathbf{Q}}_{1} - \hat{\bar{\mathbf{Q}}}_{1}   \to \bm{0}$.  
Then, we have 
\begin{equation}  \label{eq:bary4_noise_approx}
  \bar{\mathbf{y}}_4 \approx c_{\theta} \hat{\mathbf{A}}_4  \bar{\mathbf{b}} + \mathbf{w}_{4} . 
\end{equation}
Similar to Phase I, our approach yields higher received SNR than the method in \cite{ref:doubleRIS_ZhengBY_conf}, where only one user is allowed to transmit pilots to estimate the channel related to the BD-RIS 1 - BD-RIS 2 link. 
Applying the LS criterion, the vector $\bar{\mathbf{b}}$ can be estimated as 
\begin{equation} 
  \hat{\bar{\mathbf{b}}} =  c_{\theta}^{-1}  ( \hat{\mathbf{A}}_4^H \hat{\mathbf{A}}_4)^{-1} \hat{\mathbf{A}}_4^H \bar{\mathbf{y}}_4  .  
\end{equation}
The estimate of $\bar{\mathbf{B}}$ can be expressed as $\hat{\bar{\mathbf{B}}} = \operatorname{unvec}(\hat{\bar{\mathbf{b}}})$.

\subsection{Phase V: Estimation of $\bar{\mathbf{r}}_{1}$} \label{Sec:V-V}

At each time instant $t = \tau'_4 + \tau \in [\tau'_4 + 1, \tau'_4 + \tau_5]$, we design the scattering matrix $\bm{\Phi}_{2,\tau'_4 + \tau}$ of BD-RIS 2 following the idea in Theorem \ref{Theorem:rank} with the difference of replacing $\bar{\mathbf{Q}}_{1}, \bar{\mathbf{Q}}_{2}$, and $\bar{\mathbf{B}}$ with the estimates $\hat{\bar{\mathbf{Q}}}_{1}, \hat{\bar{\mathbf{Q}}}_{2}$, and $\hat{\bar{\mathbf{B}}}$, respectively. 
For the initial $ \lceil \frac{KM_1}{f}  \rceil$ time instants in this phase, i.e., for each $\tau \in [  \lceil \frac{KM_1}{f}  \rceil ]$, we set the pilot signal $\mathbf{x}_{\tau'_4 + \tau}$ and scattering matrix $\bm{\Phi}_{1,\tau'_4 + \tau}$ of BD-RIS 1 as in Section~\ref{Sec:IV-V} based on the channel estimates $\hat{\bar{\mathbf{Q}}}_{1}, \hat{\bar{\mathbf{Q}}}_{2}$, and $\hat{\bar{\mathbf{B}}}$. 
For the remaining $\tau_5 - \lceil \frac{KM_1}{f}  \rceil$ time instants, i.e., for each $\tau \in [  \lceil \frac{KM_1}{f}  \rceil + 1, \tau_5 ]$, we generate $x_{k,\tau'_4 + \tau}, \forall k$ as arbitrary unit-power scalar and the scattering matrix $\bm{\Phi}_{1,\tau'_4 + \tau}$ of BD-RIS 1 as a random unitary matrix. 
Similar to \eqref{eq:bary5r1} and \eqref{eq:bary5r1_all}, after removing the contribution made by the single BD-RIS 2 reflection link from the received signal based on the estimates $\hat{\bar{\mathbf{Q}}}_{2}$ and $\hat{\bar{\mathbf{r}}}_{2}$ and stacking the observations over $\tau_5$ time instants, the overall effective received signal 
is given by
\begin{equation} 
  \bar{\mathbf{y}}_5 \!=\! \sqrt{p} \hat{\mathbf{A}}_5 [r_{1,1,1}, \bar{\bm{r}}_{1}^T]^T  + \sqrt{p} (\mathbf{A}_5 - \hat{\mathbf{A}}_5) [r_{1,1,1}, \bar{\bm{r}}_{1}^T]^T  + \mathbf{n}_{5} + \mathbf{w}_{5}, 
\end{equation} 
where $\mathbf{A}_5$ is given in \eqref{eq:A5_2}, $\hat{\mathbf{A}}_5$ is obtained by replacing the matrices $\bar{\mathbf{Q}}_{1}, \bar{\mathbf{Q}}_{2}$, and $\bar{\mathbf{B}}$ in \eqref{eq:A5_2} with their estimates $\hat{\bar{\mathbf{Q}}}_{1}, \hat{\bar{\mathbf{Q}}}_{2}$, and $\hat{\bar{\mathbf{B}}}$, respectively, 
$\mathbf{n}_{5} = [ {\mathbf{n}}_{5, 1}^T , \ldots , {\mathbf{n}}_{5, \tau_{5}}^T ]^T$ with $\mathbf{n}_{5,\tau}= \sum_{k=1}^{K} \!\sqrt{p} x_{k,\tau'_4+\tau} \!  \sum_{m_2=1}^{M_2} (  {\bar{r}}_{2,k,m_2} \bar{\mathbf{Q}}_{2} \!-\! \hat{\bar{r}}_{2,k,m_2} \hat{\bar{\mathbf{Q}}}_{2} ) [\bm{\Phi}_{2, \tau'_4+\tau}]_{:,m_2}$ and $\hat{\bar{r}}_{2,1,1}$ obtained by substituting the estimate $\hat{\bar{\mathbf{r}}}_2$ into \eqref{eq:barr1kn11}, 
and $\mathbf{w}_{5} = [ {\mathbf{z}}_{\tau'_4+1}^T , \ldots , {\mathbf{z}}_{\tau'_4+\tau_5}^T ]^T \sim \mathcal{CN}\!\left( \bm{0}, \sigma^2\mathbf{I}_{L\tau_5}\right)$. 
By increasing the pilot lengths $\tau_{1}, \tau_{2}, \tau_{3}$, and $\tau_{4}$, the channel estimates in the previous phases approach the true values, and thus we have
\begin{equation} 
  \bar{\mathbf{y}}_5  
  \approx \sqrt{p} \hat{\mathbf{A}}_5 [r_{1,1,1}, \bar{\bm{r}}_{1}^T]^T  + \mathbf{w}_5  . \label{eq:bary5r1_noise_approx}
\end{equation}
Applying the LS criterion, the vector $\bar{\bm{r}}_{1}$ can be estimated as 
\begin{equation} 
  \hat{\bar{\bm{r}}}_{1} = ( \sqrt{p} )^{-1} \left[ (\hat{\mathbf{A}}_5^H \hat{\mathbf{A}}_5)^{-1} \right]_{2:KM_1,:} \hat{\mathbf{A}}_5 \bar{\mathbf{y}}_5   .  
\end{equation}

Following the pilot sequence design and joint configuration of two BD-RIS scattering matrices mentioned before, we can obtain the estimates $\hat{\bar{\mathbf{Q}}}_{1}, \hat{\bar{\mathbf{Q}}}_{2}, \hat{\bar{\mathbf{B}}}, \hat{\bar{\bm{r}}}_{1}$, and $\hat{\bar{\bm{r}}}_{2}$ using $T = \sum_{i = 1}^5 \tau_i$ time instants. 
Then, the estimates of the cascaded channels, denoted as $\hat{\mathbf{J}}_{1,k}, \hat{\mathbf{J}}_{2,k}$, and $\hat{\mathbf{J}}_{1,2,k}, \forall k\in[K]$, can be obtained applying \eqref{eq:J1k}, \eqref{eq:J12k}, \eqref{eq:Q1kn_2}, \eqref{eq:barr1kn11}, and \eqref{eq:Q12kmnn_2}.

\section{Numerical Results}    \label{Sec:simulation} 

In this section, we provide numerical results to verify the effectiveness of the proposed channel estimation scheme. 
The system setup is as follows.  
We assume that the BS, BD-RIS 1, and BD-RIS 2 are located at $(0,0), (15,5),$ and $(5,5)$ in meter, respectively, and all users are located in a circular region with center $(20,0)$ and radius $3$. 
The path-losses of the BD-RIS $i$ - BS link, BD-RIS $1$ - BD-RIS $2$ link, and user $k$ - BD-RIS $i$ link are modeled as 
$\beta_{{\rm R }_i, {\rm BS}} = \beta_0 d_{{\rm R }_i, {\rm BS}}^{- \alpha_{{\rm R }_i, {\rm BS}} }$, 
$\beta_{{\rm R }_1, {\rm R }_2} = \beta_0 d_{{\rm R }_1, {\rm R }_2}^{-\alpha_{{\rm R }_1, {\rm R }_2}}$, 
and $\beta_{{\rm U}_k, {\rm R }_i} = \beta_0 d_{{\rm U }_k, {\rm R }_i}^{-\alpha_{{\rm U }, {\rm R }_i}}$, respectively, $\forall k\in[K], i\in\{1,2\}$,
where $\beta_0$ measures the path-loss at the reference distance, $d_{{\rm R }_i, {\rm BS}}, d_{{\rm R }_1, {\rm R }_2}$, and $d_{{\rm U }_k, {\rm R }_i}$ denote the distances of the corresponding links, and $\alpha_{{\rm R }_i, {\rm BS}}, \alpha_{{\rm R }_1, {\rm R }_2}$, and $\alpha_{{\rm U }, {\rm R }_i}$ denote corresponding path-loss factors. 
In the numerical examples, we set $\beta_0 = -20$~dB, $\alpha_{{\rm R }_2, {\rm BS}} = \alpha_{{\rm R }_1, {\rm R }_2} = \alpha_{{\rm U }, {\rm R }_1} = 2$, and $\alpha_{{\rm R }_1, {\rm BS}} = \alpha_{{\rm U }, {\rm R }_2} = 4$, 
and model the channel between the $j$-th element of BD-RIS $i$ and the BS as $\mathbf{g}_{i,j} \sim \mathcal{CN}(\bm{0}, \beta_{{\rm R }_i, {\rm BS}}\mathbf{I}_L)$, the channel between the $m_1$-th element of BD-RIS $1$ and the $m_2$-th element of BD-RIS $2$ as $b_{m_2,m_1} \sim \mathcal{CN}(0,\beta_{{\rm R }_1, {\rm R }_2})$, and the channel between the $k$-th user and the $j$-th element of BD-RIS $i$ as $r_{i,k,j} \sim \mathcal{CN}(0,\beta_{{\rm U}_k, {\rm R }_i})$~\cite{ref:wangrui}. 
The power spectral density of the AWGN at the BS is $-169$~dBm/Hz and the channel bandwidth is $1$~MHz.  
To evaluate the channel estimation performance, we adopt the normalized mean-squared error~(NMSE) metric, which is defined as
\begin{equation}
    {\rm NMSE} = \mathbb{E} \! \left[ \tfrac{ \sum_{k=1}^K \left\| \mathbf{J}_{k} - \hat{\mathbf{J}}_{k} \right\|_F^2 
    }{ \sum_{k=1}^K \left\| \mathbf{J}_{k} \right\|_F^2 } \right], 
\end{equation}
where $\mathbf{J}_{k} = [\mathbf{J}_{1,k}, \mathbf{J}_{2,k}, \mathbf{J}_{1,2,k} ]$ and $\hat{\mathbf{J}}_{k} = [\hat{\mathbf{J}}_{1,k}, \hat{\mathbf{J}}_{2,k}, \hat{\mathbf{J}}_{1,2,k} ]$.

To the best of our knowledge, the channel estimation problem for double-BD-RIS scenarios has not been studied before. Thus, to demonstrate the performance gain of our proposed scheme, we extend the method developed in \cite{ref:BDRIS_CE_LS} for single-BD-RIS systems to double-BD-RIS cases as a benchmark. Following the idea in \cite{ref:BDRIS_CE_LS}, all entries of $\mathbf{J}_{1,k}$'s, $\mathbf{J}_{2,k}$'s, and $\mathbf{J}_{1,2,k}$'s are treated as independent elements 
and estimated applying the LS criterion based on the received signal in \eqref{eq:yt}.

\begin{figure}
    \centering
    \includegraphics[width=0.93\linewidth]{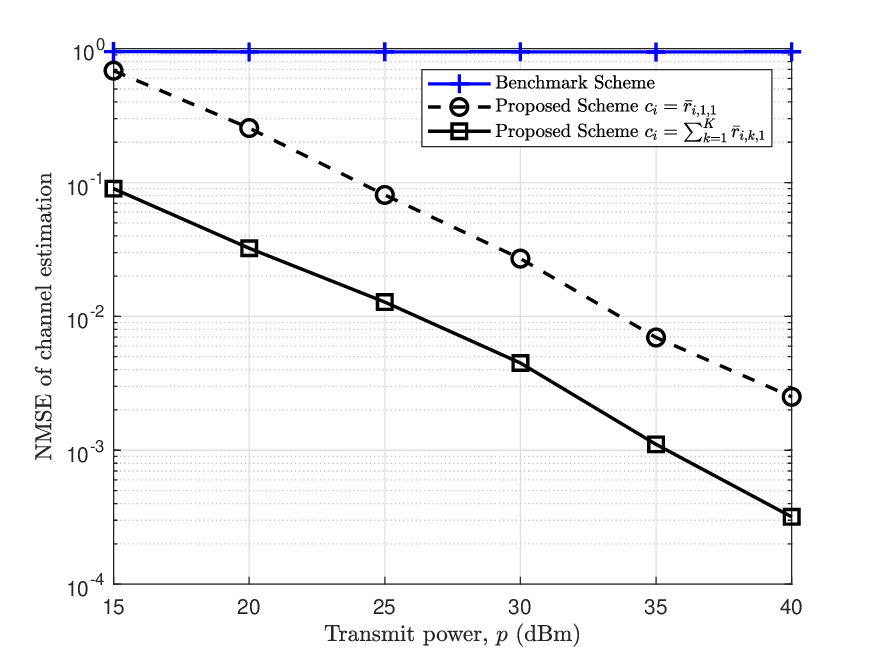}\\
    \caption{Channel estimation performance versus transmit power $p$ with $T = 64, K = 8, L = 8$, and $M_1 = M_2 = 4$. 
} \label{fig:P}  
\end{figure}

In Fig.~\ref{fig:P}, we compare the channel estimation performance achieved by the proposed scheme and the benchmark scheme under different transmit power $p$, assuming that the number of users is $K = 8$, the number of BS antennas is $L = 8$, the pilot length is $T = 64$, and the number of elements in each BD-RIS is $M_1 = M_2 = 4$. 
It is shown that the proposed scheme achieves superior performance while the benchmark scheme can hardly work. This is because the minimum required pilot length in the ideal noiseless case is $2304$ under the benchmark scheme, but reduced to $64$ (according to \eqref{eq:overhead}) under our proposed scheme. 
For the proposed scheme, channel estimation performance improves as the transmit power $p$ increases. 
Moreover, as introduced in Section~\ref{Sec:CE_AWGN}, by treating the linear combination of the channels from all users to an element of the BD-RIS and then to the BS as the reference channel and allowing all users to transmit pilots simultaneously in each phase, the proposed scheme achieves superior channel estimation performance than applying the idea in \cite{ref:wangrui,ref:doubleRIS_ZhengBY_conf}, where the scalar $c_i$ in \eqref{eq:Q1kn_2_c1} is set to be $\bar{r}_{i,1,1}, \forall i\in\{1,2\}$ and $K-1$ users are required to remain silent during the estimation of the reference channel, i.e., a typical user's channel.

\begin{figure}
    \centering
    \includegraphics[width=0.93\linewidth]{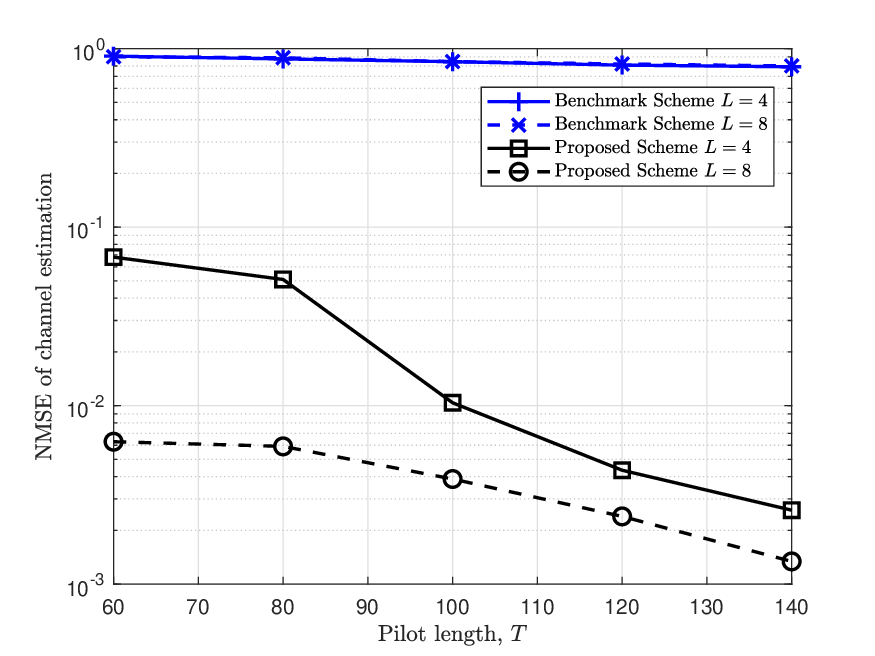}\\
    \caption{Channel estimation performance versus pilot length $T$ with $p = 30$~dBm and $K = M_1 = M_2 = 4$.
} \label{fig:T}  
\end{figure}

Fig.~\ref{fig:T} shows the performance of the proposed scheme and the benchmark scheme versus pilot length $T$ in the scenario with $p = 30$~dBm, $K = M_1 = M_2 = 4$, and $L \in \{4, 8\}$. 
The proposed scheme demonstrates a substantial performance gain over the benchmark scheme, as observed in Fig.~\ref{fig:P}. 
For the proposed scheme, the number of observations increases as $T$ increases, which effectively mitigates the effect of noise and thus contributes to a substantial reduction in channel estimation NMSE. 
Moreover, benefiting from the enlarged number of observations as shown in \eqref{eq:bary2_noise_approx}, \eqref{eq:bary4_noise_approx}, and \eqref{eq:bary5r1_noise_approx}, the NMSE is significantly reduced when we increase the number of BS antennas from $L = 4$ to $L = 8$, particularly when the pilot length $T$ is limited. 
This indicates that it is possible to compensate for the performance loss caused by insufficient pilot length by increasing the number of BS antennas.

\begin{figure}
    \centering
    \includegraphics[width=0.93\linewidth]{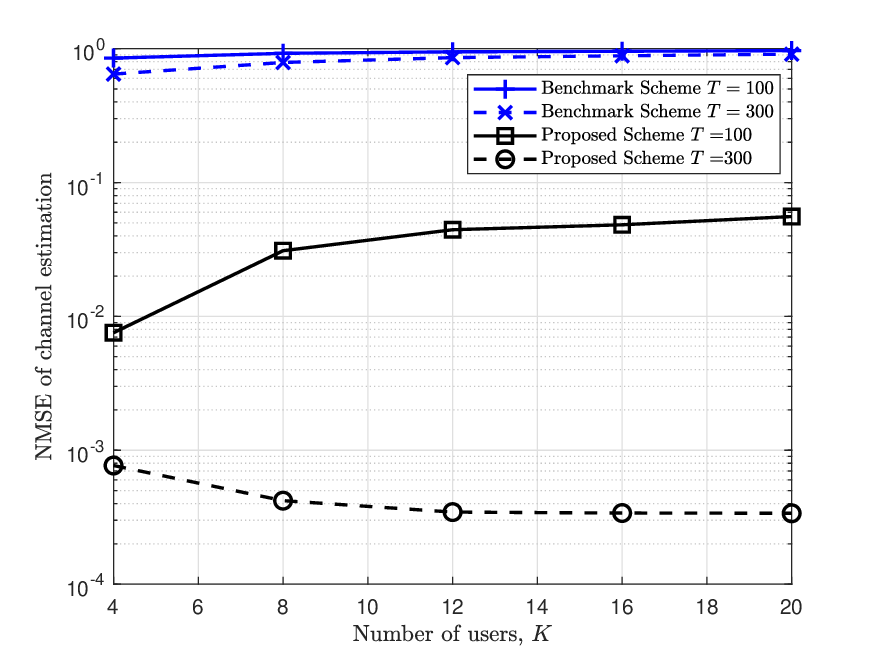}\\
    \caption{Channel estimation performance versus the number $K$ of users with $p = 30$ dBm and $L = M_1 = M_2 = 4$.
} \label{fig:K}  
\end{figure}

In Fig.~\ref{fig:K}, we present the channel estimation performance achieved by the proposed scheme and the benchmark scheme versus the number $K$ of users. Here, we assume $p = 30$ dBm, $L = M_1 = M_2 = 4$, and $T\in\{100,300\}$. 
Our proposed scheme achieves a significant NMSE performance gain compared to the benchmark scheme since much lower training overhead is required. Taking the case of $K = 20$ as an example, the benchmark scheme requires a pilot length of $5760$ to achieve perfect channel estimation in the ideal noiseless case, whereas the proposed scheme only requires $100$ time instants. 
In the case of $T = 100$, the channel estimation NMSE of the proposed scheme increases with the number $K$ of users, since the number of elements to be estimated increases at the same time and the pilot length is the bottleneck in this case. 
When $T$ is increased to $300$, 
the channel estimation NMSE shows a slight decrease as $K$ increases. This occurs because pilot length is no longer the limiting factor and the total transmit power increases with $K$, which improves the received SNR for channel estimation, 
as explained in Section~\ref{Sec:CE_AWGN}. 
Moreover, channel estimation performance is enhanced when the pilot length is increased, and this improvement is more substantial when there are a larger number of users, considering that pilot length becomes the bottleneck in this case.

\begin{figure}
    \centering
    \includegraphics[width=0.93\linewidth]{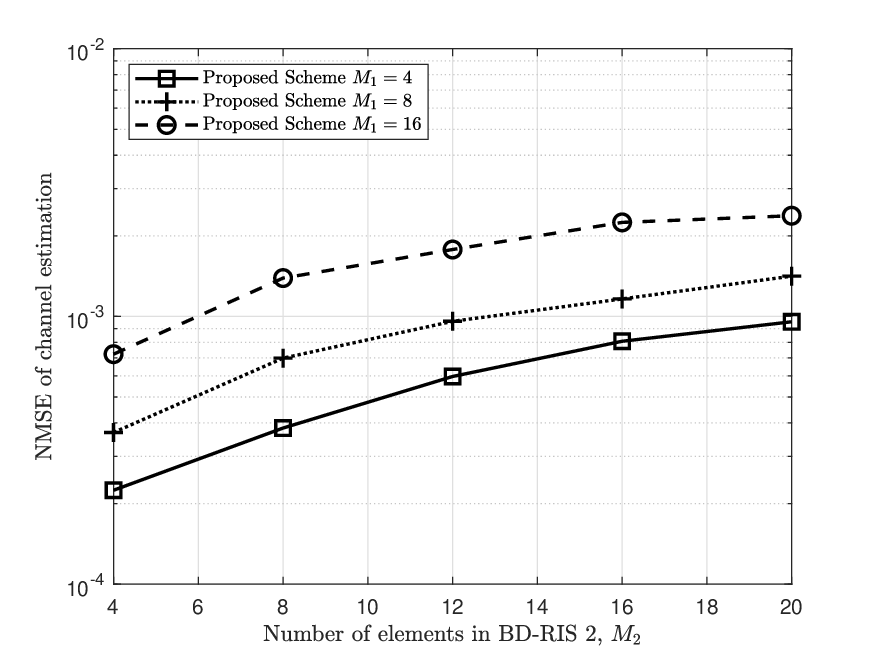}\\
    \caption{Convergence behavior versus the number $M_2$ of elements in BD-RIS 2 with $p = 30$ dBm, $T = 200$, $K = 8$, and $L = 32$. 
} \label{fig:M}  
\end{figure}

In Fig. \ref{fig:M}, we present the NMSE performance of the proposed scheme versus the number $M_2$ of elements in BD-RIS 2, assuming $p = 30$ dBm, $T = 200$, $K = 8$, $L = 32$, and $M_1 \in \{4,8,16\}$. 
A comparison with the benchmark scheme is not included since it is computation intractable suffering from the prohibitively large number of elements needed to be estimated by this scheme in the considered regime. 
As we can see, although the channel estimation accuracy degrades with the number of elements $M_1$ in BD-RIS 1 and $M_2$ in BD-RIS 2, this decline is 
not sharp due to the fully exploitation of channel correlation properties, validating the effectiveness of the proposed channel estimation framework.

\section{Conclusion} \label{Sec:conclusion} 


This paper investigated channel estimation for double-BD-RIS-aided multi-user MIMO systems, where non-diagonal scattering matrices and co-existence of single- and double-reflection links result in intricate channel coupling and a vast number of coefficients to estimate. 
To address these issues, we revealed that high-dimensional cascaded channels can be fully characterized by five low-dimensional matrices by leveraging the propagation environment overlap between single- and double-reflection links and the property that each cascaded channel is a scaled version of a reference channel. 
Building on this, we developed a channel estimation scheme to estimate these basic matrices sequentially, 
and derived closed-form overhead required for perfect channel estimation in the noise-free case. 
We proved that this overhead is on the same order as that of the double-diagonal-RIS system, and under mild conditions, is also on the same orders as that of single-BD-RIS and single-diagonal-RIS systems, implying that the substantial cooperative gain of double-BD-RIS can be realized at a channel estimation cost comparable to the other three systems. 
Then, we extended the proposed framework to practical noisy scenarios and provided extensive numerical simulations to verify its effectiveness, paving the way for realizing the full potential of multi-BD-RIS in future communication networks. 

 \appendices

 \section{Proof of Theorem \ref{Theorem:rank} }  \label{Appendix:rank}

In this appendix, we first prove that the RHS of \eqref{eq:V_B_rankF} is an upper bound of $\max_{\bm{\Phi}_2} \;\operatorname{rank}(\mathbf{F})$, and then demonstrate that this bound is achievable by designing the scattering matrix $\bm{\Phi}_2$ as in Theorem \ref{Theorem:rank}, thereby proving that the equality in \eqref{eq:V_B_rankF} holds. 

Specifically, we first construct the matrix $\mathbf{S}$ as 
\begin{equation} 
  \mathbf{S} = \begin{bmatrix}
                 \bar{\mathbf{Q}}_{1} & \bar{\mathbf{Q}}_{2} \\
                 \bar{\mathbf{B}} & -\bm{\Phi}_2^H
               \end{bmatrix}. 
\end{equation}
Applying the Guttman rank additivity formula \cite[Sec. 3.6]{ref:Matrix}, we have 
\begin{equation}  \label{eq:appendix_rankF}
  \operatorname{rank}(\mathbf{F}) = \operatorname{rank}(\mathbf{S}) - \operatorname{rank}(-\bm{\Phi}_2^H)
   = \operatorname{rank}(\mathbf{S}) - M_2.  
\end{equation}
The rank of $\mathbf{S}$ is upper-bounded by
\begin{subequations}
\begin{align}
  \operatorname{rank}(\mathbf{S}) & \leq \operatorname{rank} ([ \bar{\mathbf{Q}}_{1} , \;\! \bar{\mathbf{Q}}_{2}]) + \operatorname{rank} ([\bar{\mathbf{B}} , \;\! -\bm{\Phi}_2^H]) \\
  & = \operatorname{rank} ([ \bar{\mathbf{Q}}_{1} , \;\! \bar{\mathbf{Q}}_{2}]) + M_2 . \label{eq:appendix_rankS_bound1}
\end{align}
\end{subequations}
Likewise, we have
\begin{subequations}
\begin{align}
  \operatorname{rank}(\mathbf{S}) & \leq \operatorname{rank} \left( \begin{bmatrix}
                 \bar{\mathbf{Q}}_{1} \\
                 \bar{\mathbf{B}} 
               \end{bmatrix} \right) 
               + \operatorname{rank} \left( \begin{bmatrix}
                 \bar{\mathbf{Q}}_{2} \\
                 -\bm{\Phi}_2^H
               \end{bmatrix} \right) \\
  & = \operatorname{rank} \left( \begin{bmatrix}
                 \bar{\mathbf{Q}}_{1} \\
                 \bar{\mathbf{B}} 
               \end{bmatrix} \right)  + M_2 .  \label{eq:appendix_rankS_bound2}
\end{align}
\end{subequations}
Substituting \eqref{eq:appendix_rankS_bound1} and \eqref{eq:appendix_rankS_bound2} into \eqref{eq:appendix_rankF}, we can prove that 
\begin{equation} \label{eq:appendix_rankF_upper}
  \max_{\bm{\Phi}_2} \;\operatorname{rank}(\mathbf{F})  \leq \min \!\left\{  \operatorname{rank}\left( [ \bar{\mathbf{Q}}_{1} , \bar{\mathbf{Q}}_{2} ]  \right) ,  
  \operatorname{rank} \left( \begin{bmatrix}
                 \bar{\mathbf{Q}}_{1} \\
                 \bar{\mathbf{B}} 
               \end{bmatrix} \right) \right\}. 
\end{equation} 

In the following, we prove that this bound is achievable. Recalling that the SVD of $\bar{\mathbf{Q}}_{1}$ is $\bar{\mathbf{Q}}_{1} = \mathbf{U}_{\bar{\mathbf{Q}}_1} \bm{\Sigma}_{\bar{\mathbf{Q}}_1} \mathbf{V}_{\bar{\mathbf{Q}}_1}^H$, where $\bm{\Sigma}_{\bar{\mathbf{Q}}_{1 }} =  \operatorname{blkdiag} \{ \bm{\Lambda}_{ \bar{\mathbf{Q}}_{1 }} , \bm{0}_{L-q_1,M_1-q_1}\}$ with $\bm{\Lambda}_{ \bar{\mathbf{Q}}_{1 }} = \operatorname{diag}\{\sigma_{ \bar{\mathbf{Q}}_{1 },1} , \ldots, \sigma_{ \bar{\mathbf{Q}}_{1 },q_1} \}$. 
As introduced before, the unitary matrices are partitioned as $\mathbf{U}_{\bar{\mathbf{Q}}_1} = [\mathbf{U}_{\bar{\mathbf{Q}}_1, 1}, \mathbf{U}_{\bar{\mathbf{Q}}_1, 2}]$ and $\mathbf{V}_{\bar{\mathbf{Q}}_1} = [\mathbf{V}_{\bar{\mathbf{Q}}_1, 1}, \mathbf{V}_{\bar{\mathbf{Q}}_1, 2}]$, where $\mathbf{U}_{\bar{\mathbf{Q}}_1, 1} \in \mathbb{C}^{L \times q_1}$ and $\mathbf{V}_{\bar{\mathbf{Q}}_1, 1} \in \mathbb{C}^{M_1 \times q_1}$ consist of the singular vectors corresponding to the non-zero singular values, while $\mathbf{U}_{\bar{\mathbf{Q}}_1, 2} \in \mathbb{C}^{L \times (L-q_1)}$ and $\mathbf{V}_{\bar{\mathbf{Q}}_1, 2} \in \mathbb{C}^{M_1 \times (M_1-q_1)}$ correspond to the zero singular values. 
Accordingly, the projection matrix onto the column space of $\bar{\mathbf{Q}}_{1}$ is given by $\mathbf{P}_{\bar{\mathbf{Q}}_{1}} = \mathbf{U}_{\bar{\mathbf{Q}}_1, 1} \mathbf{U}_{\bar{\mathbf{Q}}_1, 1}^H$, and the projection matrix onto the orthogonal complement of the column space is $\mathbf{P}_{\bar{\mathbf{Q}}_{1}}^{\perp} = \mathbf{I}_L - \mathbf{P}_{\bar{\mathbf{Q}}_{1}} = \mathbf{U}_{\bar{\mathbf{Q}}_1, 2} \mathbf{U}_{\bar{\mathbf{Q}}_1, 2}^H$. The projection matrix onto the row space of $\bar{\mathbf{Q}}_{1}$ is $\bar{\mathbf{P}}_{\bar{\mathbf{Q}}_{1}} = \mathbf{V}_{\bar{\mathbf{Q}}_1, 1} \mathbf{V}_{\bar{\mathbf{Q}}_1, 1}^H$, and the projection matrix onto the null space is $\bar{\mathbf{P}}_{\bar{\mathbf{Q}}_{1}}^{\perp} = \mathbf{I}_{M_1} - \bar{\mathbf{P}}_{\bar{\mathbf{Q}}_{1}} = \mathbf{V}_{\bar{\mathbf{Q}}_1, 2} \mathbf{V}_{\bar{\mathbf{Q}}_1, 2}^H$. 
Then, $\mathbf{F}$ in \eqref{eq:V_Ft} can be rewritten as 
\begin{align}\label{eq:appendix_V_F}
  \mathbf{F} &= \bar{\mathbf{Q}}_{1} + \mathbf{P}_{\bar{\mathbf{Q}}_{1}} \bar{\mathbf{Q}}_{2}  \bm{\Phi}_2 \bar{\mathbf{B}} \bar{\mathbf{P}}_{ \bar{\mathbf{Q}}_{1}}
  + \mathbf{P}_{\bar{\mathbf{Q}}_{1}} \bar{\mathbf{Q}}_{2}  \bm{\Phi}_2 \bar{\mathbf{B}} \bar{\mathbf{P}}_{ \bar{\mathbf{Q}}_{1}}^{\perp} \notag\\
  & \;\;\;\; + \mathbf{P}_{ \bar{\mathbf{Q}}_{1}}^{\perp} \bar{\mathbf{Q}}_{2}  \bm{\Phi}_2 \bar{\mathbf{B}} \bar{\mathbf{P}}_{\bar{\mathbf{Q}}_{1}}
  + \mathbf{P}_{\bar{\mathbf{Q}}_{1}}^{\perp} \bar{\mathbf{Q}}_{2}  \bm{\Phi}_2 \bar{\mathbf{B}} \bar{\mathbf{P}}_{\bar{\mathbf{Q}}_{1}}^{\perp} . 
\end{align} 

Define the $L\times L$ diagonal matrix $\bar{\bm{\Sigma}}_{\bar{\mathbf{Q}}_{1 }} = \operatorname{blkdiag}\{ \bm{\Lambda}_{\bar{\mathbf{Q}}_{1 }}, \mathbf{I}_{L-q_1} \}$. 
Considering that we have $\operatorname{rank}(\mathbf{AB}) = \operatorname{rank}(\mathbf{B})$ if $\mathbf{A}$ has full column rank, and have $\operatorname{rank}(\mathbf{AB}) = \operatorname{rank}(\mathbf{A})$ if $\mathbf{B}$ has full row rank, we can obtain that $\operatorname{rank}(\mathbf{F}) = \operatorname{rank}(\breve{\mathbf{F}})$ with $\breve{\mathbf{F}}$ given by 
\begin{subequations}
\begin{align}
  & \breve{\mathbf{F}} = \bar{\bm{\Sigma}}_{\bar{\mathbf{Q}}_{1 }}^{-1} \mathbf{U}_{\bar{\mathbf{Q}}_1}^H \mathbf{F} \mathbf{V}_{\bar{\mathbf{Q}}_1} \\
  & = \operatorname{blkdiag}\{ \mathbf{I}_{q_1}, \bm{0}_{ L-q_1 , M_1-q_1} \}  \notag\\
  & \;\;\;\; +  \begin{bmatrix} \bm{\Lambda}_{\bar{\mathbf{Q}}_{1 }}^{-1} \mathbf{U}_{\bar{\mathbf{Q}}_1,1}^H \\ \bm{0}_{ L-q_1, L} \end{bmatrix}
  \bar{\mathbf{Q}}_{2}  \bm{\Phi}_2 \bar{\mathbf{B}} [ \mathbf{V}_{\bar{\mathbf{Q}}_1,1}, \bm{0}_{ M_1, M_1-q_1 } ] \notag\\
  & \;\;\;\; +  \begin{bmatrix} \bm{\Lambda}_{\bar{\mathbf{Q}}_{1 }}^{-1} \mathbf{U}_{\bar{\mathbf{Q}}_1,1}^H \\ \bm{0}_{ L-q_1, L} \end{bmatrix} \bar{\mathbf{Q}}_{2}  \bm{\Phi}_2 \bar{\mathbf{B}} [ \bm{0}_{ M_1 , q_1 } , \mathbf{V}_{ \bar{\mathbf{Q}}_1,2}   ]  \notag\\
  & \;\;\;\; + \begin{bmatrix} \bm{0}_{q_1, L}  \\ \mathbf{U}_{ \bar{\mathbf{Q}}_1,2}^H  \end{bmatrix} \bar{\mathbf{Q}}_{2}  \bm{\Phi}_2 \bar{\mathbf{B}} [ \mathbf{V}_{\bar{\mathbf{Q}}_1,1}, \bm{0}_{ M_1, M_1-q_1  } ] \notag\\
  & \;\;\;\; + \begin{bmatrix} \bm{0}_{q_1, L}  \\ \mathbf{U}_{\bar{\mathbf{Q}}_1,2}^H  \end{bmatrix}  \bar{\mathbf{Q}}_{2}  \bm{\Phi}_2 \bar{\mathbf{B}} [ \bm{0}_{ M_1, q_1 } , \mathbf{V}_{\bar{\mathbf{Q}}_1,2}   ] \\
  & = \!\begin{bmatrix}  
  \mathbf{I}_{q_1} \!\! + \! \bm{\Lambda}_{ \bar{\mathbf{Q}}_{1 }}^{ -1}  \mathbf{U}_{ \bar{\mathbf{Q}}_1 ,1}^H  \bar{\mathbf{Q}}_{2}   \bm{\Phi}_2 \bar{\mathbf{B}}  \mathbf{V}_{ \bar{\mathbf{Q}}_1 ,1}  \!\!   & \bm{\Lambda}_{ \bar{\mathbf{Q}}_{1 }}^{ -1}  \mathbf{U}_{ \bar{\mathbf{Q}}_1 ,1}^H  \bar{\mathbf{Q}}_{2}   \bm{\Phi}_2 \tilde{\mathbf{B}}   \\
  \tilde{\mathbf{Q}}_{2}  \bm{\Phi}_2 \bar{\mathbf{B}} \mathbf{V}_{\bar{\mathbf{Q}}_1,1}  \!\!   & \tilde{\mathbf{Q}}_{2}  \bm{\Phi}_2 \tilde{\mathbf{B}}   
  \end{bmatrix} \!,  \label{eq:appendix_V_Ftilde}
\end{align}
\end{subequations}
where the matrices $\tilde{\mathbf{Q}}_{2}$ and $\tilde{\mathbf{B}}$ are defined in Theorem \ref{Theorem:rank}. 
The SVD of $\tilde{\mathbf{Q}}_{2}$ is expressed as $\tilde{\mathbf{Q}}_{2} = \mathbf{U}_{\tilde{\mathbf{Q}}_2} \bm{\Sigma}_{\tilde{\mathbf{Q}}_2} \mathbf{V}_{\tilde{\mathbf{Q}}_2}^H$, where $\mathbf{U}_{\tilde{\mathbf{Q}}_2} \in \mathbb{C}^{(L-q_1) \times (L-q_1)}$ and $\mathbf{V}_{\tilde{\mathbf{Q}}_2} \in \mathbb{C}^{M_2\times M_2}$ are unitary matrices, and $\bm{\Sigma}_{\tilde{\mathbf{Q}}_2}$ is an $(L-q_1) \times M_2$ rectangular diagonal matrix containing $\tilde{q}_2$ nonzero singular values. 
The SVD of $\tilde{\mathbf{B}}$ is expressed as $\tilde{\mathbf{B}} = \mathbf{U}_{\tilde{\mathbf{B}} } \bm{\Sigma}_{\tilde{\mathbf{B}} } \mathbf{V}_{\tilde{\mathbf{B}} }^H$, where $\mathbf{U}_{\tilde{\mathbf{B}}} \in \mathbb{C}^{M_2\times M_2}$ and $\mathbf{V}_{\tilde{\mathbf{B}} } \in \mathbb{C}^{(M_1-q_1) \times (M_1-q_1)}$ are unitary matrices, and $\bm{\Sigma}_{\tilde{\mathbf{B}}}$ is an $M_2\times (M_1-q_1)$ rectangular diagonal matrix containing $\tilde{b}$ nonzero singular values. 
The scattering matrix $\bm{\Phi}_2$ in \eqref{eq:appendix_V_Ftilde} is designed as in \eqref{eq:V_B_phi}.

Then, we have $ \operatorname{rank}(\breve{\mathbf{F}}) = \operatorname{rank}(\bar{\mathbf{F}}) $ with $\bar{\mathbf{F}}$ given by 
\begin{subequations}
\begin{align}
  \bar{\mathbf{F}} \!& = \!\operatorname{blkdiag}  \{ \mathbf{I}_{q_1} \!, \! e^{-\jmath \phi} \mathbf{U}_{\!\tilde{\mathbf{Q}}_2}^H \}
  \!\operatorname{rank}(\breve{\mathbf{F}}) 
  \operatorname{blkdiag} \{ \mathbf{I}_{q_1} \!, \!\! \mathbf{V}_{\!\tilde{\mathbf{B}}}  \} \\
  & = \begin{bmatrix}  \bar{\mathbf{F}}_{11} & \bar{\mathbf{F}}_{12} \\
  \bar{\mathbf{F}}_{21} & \bar{\mathbf{F}}_{22}  \end{bmatrix}, 
\end{align}
\end{subequations}
where 
\begin{equation}\label{eq:appendix_V_Fbar11}
  \bar{\mathbf{F}}_{11} = \mathbf{I}_{q_1} + e^{\jmath \phi} \bm{\Lambda}_{ \bar{\mathbf{Q}}_{1 }}^{ -1}  \mathbf{U}_{ \bar{\mathbf{Q}}_1 ,1}^H  \bar{\mathbf{Q}}_{2}   \mathbf{V}_{\tilde{\mathbf{Q}}_2}  \mathbf{U}_{\tilde{\mathbf{B}} }^H   \bar{\mathbf{B}}  \mathbf{V}_{ \bar{\mathbf{Q}}_1 ,1}, 
\end{equation}
\begin{equation}\label{eq:appendix_V_Fbar12}
  \bar{\mathbf{F}}_{12} = e^{\jmath \phi} \bm{\Lambda}_{ \bar{\mathbf{Q}}_{1 }}^{ -1}  \mathbf{U}_{ \bar{\mathbf{Q}}_1 ,1}^H  \bar{\mathbf{Q}}_{2}   \mathbf{V}_{\tilde{\mathbf{Q}}_2}  \bm{\Sigma}_{\tilde{\mathbf{B}} } , 
\end{equation}
\begin{equation}\label{eq:appendix_V_Fbar21}
  \bar{\mathbf{F}}_{21} =   \bm{\Sigma}_{\tilde{\mathbf{Q}}_2}  \mathbf{U}_{\tilde{\mathbf{B}} }^H  \bar{\mathbf{B}} \mathbf{V}_{\bar{\mathbf{Q}}_1,1}, 
\end{equation}
\begin{equation}\label{eq:appendix_V_Fbar22}
  \bar{\mathbf{F}}_{22} = \bm{\Sigma}_{\tilde{\mathbf{Q}}_2}  \bm{\Sigma}_{\tilde{\mathbf{B}} } .  
\end{equation}
We can obtain that 
\begin{subequations}
\begin{align}
  & \operatorname{rank}(\bar{\mathbf{F}}) \notag\\
  & \geq \operatorname{rank} \left( \begin{bmatrix}  \bar{\mathbf{F}}_{11} & [\bar{\mathbf{F}}_{12}]_{:,1:c } \\
  [\bar{\mathbf{F}}_{21}]_{1:c,:} & [\bar{\mathbf{F}}_{22}]_{1:c,1:c}  \end{bmatrix} \right) \\
  & = \operatorname{rank}([\bar{\mathbf{F}}_{22}]_{1:c,1:c}) \notag\\
  & \;\;\; + \operatorname{rank}( \bar{\mathbf{F}}_{11} - [\bar{\mathbf{F}}_{12}]_{:,1:c } ([\bar{\mathbf{F}}_{22}]_{1:c,1:c})^{-1} [\bar{\mathbf{F}}_{21}]_{1:c,:} ), \label{eq:appendix_V_Fbar_rank}
\end{align}
\end{subequations}
where 
\begin{equation}
  c = \operatorname{rank}([\bar{\mathbf{F}}_{22}]_{1:c,1:c}) = \min\{\tilde{q}_2, \tilde{b}\} ,  
\end{equation}
with $\tilde{q}_2$ and $\tilde{b}$ denoting the ranks of $\tilde{\mathbf{Q}}_{2}$ and $\tilde{\mathbf{B}}$, respectively. 
The matrix in the second term of the RHS of \eqref{eq:appendix_V_Fbar_rank} satisfies
\begin{equation}
  \bar{\mathbf{F}}_{11} - [\bar{\mathbf{F}}_{12}]_{:,1:c } ([\bar{\mathbf{F}}_{22}]_{1:c,1:c})^{-1} [\bar{\mathbf{F}}_{21}]_{1:c,:} 
  =  \mathbf{I}_{q_1} + e^{\jmath \phi}   \tilde{\mathbf{F}}, 
\end{equation}
where the expression of the matrix $\tilde{\mathbf{F}}\in \mathbb{C}^{q_1\times q_1}$ is given in \eqref{eq:V_Ftilde}. 
Denote the eigenvalues of $\tilde{\mathbf{F}}$ as $\lambda_{\tilde{\mathbf{F}}, 1}, \ldots, \lambda_{\tilde{\mathbf{F}}, q_1 }$. Then, the eigenvalues of the matrix $\mathbf{I}_{ q_1 } + e^{\jmath \phi} \tilde{\mathbf{F}} $ are denoted as $1 + e^{\jmath \phi} \lambda_{\tilde{\mathbf{F}}, 1}, \ldots, 1  + e^{\jmath \phi} \lambda_{\tilde{\mathbf{F}}, q_1} $. 
Define the set $\tilde{\Theta}$ as in \eqref{eq:V_B_phi_Thetatilde}. 
Then, we choose the phase $\phi$ satisfying that
\begin{equation}
  \phi \in [0, 2\pi) \setminus \tilde{\Theta} .  
\end{equation}
This condition guarantees that $\operatorname{rank}\left( \mathbf{I}_{ q_1 } + e^{\jmath \phi} \tilde{\mathbf{F}}  \right) = q_1$. 
As a result, we can obtain that 
\begin{equation} \label{eq:appendix_rankF_lower1}
  \max_{\bm{\Phi}_2} \;\operatorname{rank}(\mathbf{F})  
  \geq q_1 + \min\{\tilde{q}_2, \tilde{b}\}. 
\end{equation}

It is satisfied that 
\begin{subequations}
\begin{align} 
  \tilde{q}_2 & = \operatorname{rank}(\tilde{\mathbf{Q}}_{2}) \label{eq:appendix_tildeq2_1} \\
  & = \operatorname{rank}(  \mathbf{U}_{\bar{\mathbf{Q}}_{1},2}^H  {\mathbf{P}}_{\bar{\mathbf{Q}}_{1}}^{\perp}  \bar{\mathbf{Q}}_{2} ) \label{eq:appendix_tildeq2_2} \\
  & = \operatorname{rank}( \mathbf{U}_{\bar{\mathbf{Q}}_{1}}^H  {\mathbf{P}}_{\bar{\mathbf{Q}}_{1}}^{\perp} \bar{\mathbf{Q}}_{2} ) \label{eq:appendix_tildeq2_3} \\
  & = \operatorname{rank}(  {\mathbf{P}}_{\bar{\mathbf{Q}}_{1}}^{\perp} \bar{\mathbf{Q}}_{2} ) \label{eq:appendix_tildeq2_4} \\
  & = \operatorname{rank} ([ \bar{\mathbf{Q}}_{1} , \;\! \bar{\mathbf{Q}}_{2}]) - \operatorname{rank} ( \bar{\mathbf{Q}}_{1} ), \label{eq:appendix_tildeq2_5} 
\end{align}
\end{subequations}
where \eqref{eq:appendix_tildeq2_2} follows because $\tilde{\mathbf{Q}}_{2} = \mathbf{U}_{\bar{\mathbf{Q}}_{1},2}^H  \bar{\mathbf{Q}}_{2} = \mathbf{U}_{\bar{\mathbf{Q}}_{1},2}^H  {\mathbf{P}}_{\bar{\mathbf{Q}}_{1}}^{\perp}  \bar{\mathbf{Q}}_{2}$, 
\eqref{eq:appendix_tildeq2_3} follows because $\mathbf{U}_{\bar{\mathbf{Q}}_{1}}^H  {\mathbf{P}}_{\bar{\mathbf{Q}}_{1}}^{\perp} \bar{\mathbf{Q}}_{2}$ is equal to $\begin{bmatrix} \bm{0}_{q_1, M_2} \\ \mathbf{U}_{\bar{\mathbf{Q}}_{1},2}^H  {\mathbf{P}}_{\bar{\mathbf{Q}}_{1}}^{\perp}  \bar{\mathbf{Q}}_{2} \end{bmatrix}$ and it has the same rank as $\mathbf{U}_{\bar{\mathbf{Q}}_{1},2}^H  {\mathbf{P}}_{\bar{\mathbf{Q}}_{1}}^{\perp}  \bar{\mathbf{Q}}_{2}$, 
and \eqref{eq:appendix_tildeq2_5} follows because $\operatorname{rank}([\mathbf{A}, \mathbf{B}]) = \operatorname{rank}(\mathbf{A}) + \operatorname{rank}( {\mathbf{P}}_{\mathbf{A}}^{\perp} \mathbf{B} )$ for any matrices $\mathbf{A}$ and $\mathbf{B}$ with the same number of rows \cite[Sec. 3.6]{ref:Matrix}. 
Likewise, we have
\begin{subequations}
\begin{align} 
  \tilde{b} & = \operatorname{rank}(\tilde{\mathbf{B}}) \label{eq:appendix_tildeb_1} \\
  & = \operatorname{rank}( \bar{\mathbf{B}} \bar{\mathbf{P}}_{\bar{\mathbf{Q}}_{1}}^{\perp} \mathbf{V}_{\bar{\mathbf{Q}}_{1},2}  ) \label{eq:appendix_tildeb_2} \\
  & = \operatorname{rank}( \bar{\mathbf{B}} \bar{\mathbf{P}}_{\bar{\mathbf{Q}}_{1}}^{\perp} \mathbf{V}_{\bar{\mathbf{Q}}_{1}}  ) \label{eq:appendix_tildeb_3} \\
  & = \operatorname{rank}( \bar{\mathbf{B}} \bar{\mathbf{P}}_{\bar{\mathbf{Q}}_{1}}^{\perp} ) \label{eq:appendix_tildeb_4} \\
  & = \operatorname{rank} \left( \begin{bmatrix}
                 \bar{\mathbf{Q}}_{1} \\
                 \bar{\mathbf{B}} 
               \end{bmatrix} \right) - \operatorname{rank} ( \bar{\mathbf{Q}}_{1} ) . \label{eq:appendix_tildeb_5} 
\end{align} 
\end{subequations}
Substituting \eqref{eq:appendix_tildeq2_5} and \eqref{eq:appendix_tildeb_5} into \eqref{eq:appendix_rankF_lower1}, we have 
\begin{equation} \label{eq:appendix_rankF_lower}
  \max_{\bm{\Phi}_2} \;\operatorname{rank}(\mathbf{F})  
  \geq \min \!\left\{  \operatorname{rank}\left( [ \bar{\mathbf{Q}}_{1} , \bar{\mathbf{Q}}_{2} ]  \right) ,  
  \operatorname{rank} \left( \begin{bmatrix}
                 \bar{\mathbf{Q}}_{1} \\
                 \bar{\mathbf{B}} 
               \end{bmatrix} \right) \right\}. 
\end{equation} 
Together with \eqref{eq:appendix_rankF_upper}, we can prove \eqref{eq:V_B_rankF} in Theorem \ref{Theorem:rank}.


 \section{Proof of Corollary \ref{Corollary:rank} }  \label{Appendix:rank2}

We can obtain that 
\begin{subequations}
\begin{align} 
  \operatorname{rank}\left( [ \bar{\mathbf{Q}}_{1}, \bar{\mathbf{Q}}_{2}] \right)  
  & = \operatorname{rank}\left( \bar{\mathbf{Q}}_{1} \right) + \operatorname{rank}\left( \mathbf{P}_{\bar{\mathbf{Q}}_{1}}^{\perp} \bar{\mathbf{Q}}_{2} \right) \label{eq:appendix_rankQ1Q2_1}\\
  & \leq \operatorname{rank}( \bar{\mathbf{Q}}_{1}) + \operatorname{rank}(\bar{\mathbf{Q}}_{2})  \label{eq:appendix_rankQ1Q2_2}\\
  & = q_1 + q_2,  \label{eq:appendix_rankQ1Q2_3}
\end{align}
\end{subequations}
where \eqref{eq:appendix_rankQ1Q2_1} follows from \cite[Sec. 3.6]{ref:Matrix}, and the equality in \eqref{eq:appendix_rankQ1Q2_2} holds if and only if the intersection of the column spaces of $\bar{\mathbf{Q}}_{1}$ and $\bar{\mathbf{Q}}_{2}$ is trivial, 
i.e., $\dim (\operatorname{Col}( \bar{\mathbf{Q}}_{1}) \cap \operatorname{Col}( \bar{\mathbf{Q}}_{2}) )= 0$. 
Likewise, we have
\begin{subequations}
\begin{align} 
  \operatorname{rank}\left( \begin{bmatrix}
                 \bar{\mathbf{Q}}_{1} \\
                 \bar{\mathbf{B}} 
               \end{bmatrix} \right)  
  & = \operatorname{rank}\left( \bar{\mathbf{Q}}_{1} \right) + \operatorname{rank}\left( \bar{\mathbf{B}} \bar{\mathbf{P}}_{ \bar{\mathbf{Q}}_1}^{\perp}  \right) \label{eq:appendix_rankQ1B_1}\\
  & \leq \operatorname{rank}( \bar{\mathbf{Q}}_{1}) +  \operatorname{rank}( \bar{\mathbf{B}} )  \label{eq:appendix_rankQ1B_2}\\
  & = q_1 + b. \label{eq:appendix_rankQ1B_3}
\end{align}
\end{subequations}
The equality in \eqref{eq:appendix_rankQ1B_2} holds if and only if the intersection of the row spaces of $\bar{\mathbf{Q}}_{1}$ and $\bar{\mathbf{B}}$ is trivial, 
i.e., $\dim (\operatorname{Row}(\bar{\mathbf{Q}}_{1}) \cap \operatorname{Row}(\bar{\mathbf{B}}) )= 0$. 
Substituting \eqref{eq:appendix_rankQ1Q2_3} and \eqref{eq:appendix_rankQ1B_3} into \eqref{eq:V_B_rankF} and considering that the rank of any matrix is upper-bounded by its dimensions, we can obtain \eqref{eq:V_B_rankF_upper}.

Since the rank of a concatenated matrix is at least the maximum rank of its sub-blocks, we have
\begin{subequations}
\begin{align}
  \operatorname{rank}\left( [ \bar{\mathbf{Q}}_{1}, \bar{\mathbf{Q}}_{2}] \right) 
  & \geq \max\{ \operatorname{rank}( \bar{\mathbf{Q}}_{1}), \operatorname{rank}( \bar{\mathbf{Q}}_{2}) \} \label{eq:appendix_rankQ1Q2_lower1} \\
  & = \max\{ q_1, q_2 \}. \label{eq:appendix_rankQ1Q2_lower2}
\end{align}
\end{subequations}
The lower bound corresponds to the worst-case scenario where the channel subspaces are aligned as much as possible. Specifically, when $\operatorname{Col}( \bar{\mathbf{Q}}_{2}) \subseteq \operatorname{Col}( \bar{\mathbf{Q}}_{1})$, the equality $\operatorname{rank}\left( [ \bar{\mathbf{Q}}_{1}, \bar{\mathbf{Q}}_{2}] \right) = q_1$ holds. 
When $\operatorname{Col}( \bar{\mathbf{Q}}_{1}) \subseteq \operatorname{Col}( \bar{\mathbf{Q}}_{2})$, the equality $\operatorname{rank}\left( [ \bar{\mathbf{Q}}_{1}, \bar{\mathbf{Q}}_{2}] \right) = q_2$ holds. 
Likewise, we have
\begin{subequations}
\begin{align}
  \operatorname{rank}\left( \begin{bmatrix} \bar{\mathbf{Q}}_{1} \\ \bar{\mathbf{B}} \end{bmatrix} \right) 
  & \geq \max\{ \operatorname{rank}( \bar{\mathbf{Q}}_{1}), \operatorname{rank}(\bar{\mathbf{B}}) \} \label{eq:appendix_rankQ1B_lower1} \\
  & = \max\{ q_1, b \}. \label{eq:appendix_rankQ1B_lower2}
\end{align}
\end{subequations}
When $\operatorname{Row}(\bar{\mathbf{B}}) \subseteq \operatorname{Row}(\bar{\mathbf{Q}}_{1})$, the equality $\operatorname{rank}\left( \begin{bmatrix} \bar{\mathbf{Q}}_{1} \\ \bar{\mathbf{B}} \end{bmatrix} \right) = q_1$ holds. 
When $\operatorname{Row}( \bar{\mathbf{Q}}_{1}) \subseteq \operatorname{Row}(\bar{\mathbf{B}})$, the equality $\operatorname{rank}\left( \begin{bmatrix} \bar{\mathbf{Q}}_{1} \\ \bar{\mathbf{B}} \end{bmatrix} \right) = b$ holds. 
Substituting \eqref{eq:appendix_rankQ1Q2_lower2} and \eqref{eq:appendix_rankQ1B_lower2} into \eqref{eq:V_B_rankF}, we can obtain \eqref{eq:V_B_rankF_lower}.

\bibliographystyle{IEEEtran}  
\bibliography{IEEEabrv,reference}

\end{document}